\documentclass[lineno]{JFM-FLM_Au}

\usepackage{multirow}
\usepackage[table]{xcolor}
\usepackage{booktabs}

\lefttitle{P. Patra, D.L. Koch and A. Roy}
\righttitle{Journal of Fluid Mechanics}

\title{Collision efficiency of rapidly settling particle pairs in a turbulent flow}
\author{Pijush Patra\aff{1}\footnote{Present address: Nordita, KTH Royal Institute of Technology and Stockholm University, Stockholm 10691, Sweden},
Donald L. Koch\aff{2}
\and Anubhab Roy\aff{1}}
\affiliation{\aff{1}Department of Applied Mechanics, Indian Institute of Technology Madras, Chennai, Tamil Nadu 600036, India
\aff{2}Smith School of Chemical and Biomolecular Engineering, Cornell University,
Ithaca, New York 14853, USA}
\corresau{Anubhab Roy, \email{anubhab@iitm.ac.in}}

\begin{document}
\maketitle
\begin{abstract}
We investigate the collision dynamics of hydrodynamically interacting inertialess spherical particle pairs sedimenting in a homogeneous isotropic turbulent flow. The analysis focuses on the rapid-settling limit, in which the particle settling time across a Kolmogorov eddy is much shorter than the Kolmogorov time scale. We also consider continuum breakdown during lubrication interactions, which is important when the separation the particles is comparable to the $O(100)$ nm mean-free path of a gaseous media. Owing to the sub-Kolmogorov particle sizes considered here, we approximate the local flow field in the vicinity of a particle pair as a stochastic linear flow induced by the background turbulence. In the rapid-settling regime, the cumulative effect of turbulent strain fluctuations is weak, and the relative particle motion may therefore be described as a diffusive process. In addition, hydrodynamic interactions generate a net relative drift between the particle pairs. We obtain the hydrodynamic diffusivity and relative drift velocity from the Lagrangian autocorrelation function of the fluid velocity gradient evaluated along the settling trajectory. The rapid-settling assumption further enables us to relate the autocorrelation function to the turbulence energy spectrum. Using these results, we solve the advection-diffusion equation for the pair probability density function to determine the collision rate. We show that the ideal collision rate increases monotonically with increasing relative strength of gravity to turbulence, whereas the collision efficiency decreases monotonically over the same range.
\end{abstract}

\begin{keywords}
Authors should not enter keywords on the manuscript, as these must be chosen by the author during the online submission process and will then be added during the typesetting process (see \href{https://www.cambridge.org/core/journals/journal-of-fluid-mechanics/information/list-of-keywords}{Keyword PDF} for the full list).  Other classifications will be added at the same time.
\end{keywords}

\section{Introduction}\label{sec: Introduction}
Collision of particles sedimenting in a turbulent flow is relevant to many environmental and industrial processes. Applications include the growth of water droplets in warm clouds and the aggregation of soot particles during industrial emissions. Analysis of coagulation finds application in industrial settings also, such as carbon black aggregation in aerosol reactors \citep{buesser2012design}. The evolution of the particle size distribution in such systems crucially depends on the collision rate between the particles. The combined effect of turbulence, gravity, and interparticle interactions drives the collision dynamics in these systems. A study of this problem might explain the \textit{condensation-coalescence bottleneck} (or the `size gap' of $15-40$ \textmu m droplets) in warm rain formation, where neither condensation nor gravitational collision alone is the dominant growth mechanism. An accurate estimation of the collisional growth rate of cloud droplets in the `size gap' regime is crucial for obtaining a physically correct rain-forming drop size distribution (DSD) that sets the time of rain formation in warm clouds and plays a vital role in the atmospheric thermal budget (see \citealt{slingo1990sensitivity,feingold1999impact,peng2002cloud}). Thus a detailed study of the collision process can aid in analyzing the growth mechanism of cloud droplets and improve the parameterizations of cloud microphysical processes. Better parameterizations can, in turn, significantly reduce the uncertainties involved in weather forecasting and climate models. The present study will focus on collisions of inertialess particle pairs settling rapidly in a turbulent atmosphere while interacting through non-continuum hydrodynamics at close separations. 

Particle collision in turbulence was first studied by \cite{saffman1956collision}, who calculated the collision rate between two non-interacting spherical particles in the `persistent' strain limit. Therefore, they approximated the turbulent flow field experienced by the sub-Kolmogorov particles as a pseudo-steady uniaxial compressional flow and found the collision rate to be $(8 \pi /15)^{1/2} n_1 n_2 \varGamma_{\eta} (a_1+a_2)^3$, where $a_1$ and $a_2$ are the radii of the two spheres with number density $n_1$ and $n_2$ respectively, $\varGamma_{\eta}=(\epsilon/\nu_f)^{1/2}$ is the Kolmogorov shear rate, $\epsilon$ is the turbulent dissipation rate and $\nu_f$ is the kinematic viscosity of the surrounding fluid. In the limit of small total strain (i.e., the product of the characteristic strain rate and its correlation time is small), \cite{brunk1997hydrodynamic} considered an inertialess particle pair subjected to an isotropic random velocity field and derived a pair probability equation containing a diffusion tensor and a relative drift. They calculated the drift-diffusion fluxes using the fluid Lagrangian velocity-gradient autocorrelation function and found the coagulation rate constant for non-interacting particle pairs to be $(4 \pi /5)n_1n_2(\varGamma_{\eta} \tau_S) \varGamma_{\eta} (a_1+a_2)^3$, where $\tau_S$ is the strain rate correlation time. This functional form is similar to the form obtained by \cite{saffman1956collision} except for the extra dependency on the total strain rate $\varGamma_{\eta}\tau_S$. In a subsequent study, \cite{brunk1998turbulent} developed a stochastic simulation technique by assuming Gaussian statistics for the fluctuating velocity gradient to compute the turbulent coagulation rate for an arbitrary total strain with and without particle interactions. For a specific case of $\tau_S \varGamma_{\eta} = 2.3$ representing isotropic turbulence as estimated from direct numerical simulation (DNS), they found the normalized coagulation rate constant for two non-interacting equal-sized particles to be $8.62 \pm 0.02$. On the other hand, \cite{von1917investigation} found the collision rate for two non-interacting spheres settling in a quiescent fluid to be $2\pi n_1 n_2(a_1 + a_2)^2 \rho_pg\left( a_1^2-a_2^2\right)/\left(9\mu_f\right)$, where $g$ is the magnitude of the acceleration due to gravity $\boldsymbol{g}$, $\rho_p$ is the density of the particles and $\mu_f$ is the dynamic viscosity. \cite{davis1984rate} calculated the collision efficiency of a dilute polydisperse system of sedimenting spheres with hydrodynamic and van der Waals interactions. For the mixed problem (gravity coupled with turbulence), \cite{li2018effect} performed DNS to determine the collision rate in the absence of interparticle interactions only for a few relative strengths of gravity to turbulence. Also, the maximum value of the Taylor microscale Reynolds ($Re_{\lambda}\equiv (15u'^4/(\nu_f\epsilon))^{1/2}$, with $u'$ being the root mean square (r.m.s.) velocity) in their calculation was 158, whereas, in clouds, it is about $O(10^4)$. \citet{dhanasekaran2021collision}, \citet{dhanasekaran2021turbulent}, and \citet{patra_koch_roy_2022} have respectively calculated the collision rate of bidisperse spheres settling in a uniaxial compressional flow, a turbulent flow, and a simple shear flow for a wide range of the parameter representing the relative strengths of gravity to the respective background flow. While \citet{dhanasekaran2021turbulent} assumed that the particle pairs sampled the turbulence in a Lagrangian reference frame, the present study addresses the collision dynamics of particle pairs that settle rapidly through the turbulent eddies.

Many natural and industrial systems, such as atmospheric clouds (see \citealt{Grabowski2013growth}), aerosol reactors (see \citealt{Balthasar2002detailed}), and separators, have low particle volume fractions (about $O(10^{-6})$). So, we consider a dilute system and restrict our analysis to binary interactions of particles with radii $a_1$ and $a_2$. We ignore interface mobility and assume the drops to be rigid spheres. This assumption is valid for small water droplets (radii less than 25 \textmu m) due to their high drop-to-medium viscosity ratio. The interparticle interactions, particularly hydrodynamic interactions, significantly influence the collision rate. At close separations, the continuum assumption of hydrodynamic interactions fails, and near-field non-continuum interactions become the dominant mechanism of collisions in media with long mean free paths (see \citealt{dhanasekaran2021collision,dhanasekaran2021turbulent,PhysRevFluids.7.064308}). In aqueous suspensions, van der Waals forces overcome the continuum lubrication forces and allow surface-to-surface contact between particle pairs in a finite time (see  \citealt{batchelor1982sedimentation1,davis1984rate,wang1994collision}). However, the importance of non-continuum lubrication interactions on collisions in gaseous media has received less attention in previous studies. More importantly, for particles in the radius range of $15-40$ \textmu m (i.e., the `size gap' regime), the non-continuum effect dominates over the effects of interfacial mobility, compressibility, deformability, and van der Waals force (see figure 2 of \citealt{dhanasekaran2021collision}). The Knudsen number defined as $Kn=\lambda_g/a^*$ quantifies the strength of non-continuum effects. Here, $\lambda_g$ is the mean free path of the medium ($\lambda_g \approx 100$ nm for air at typical cloud pressure), and $a^*=(a_1+a_2)/2$ is the average radius of the two interacting spheres. \cite{hocking1973effect} initially tried to capture the non-continuum effects through Maxwell slip boundary conditions. Using Hocking's results, \cite{davis1984rate} calculated the role of non-continuum effects on the collision rate for two differentially settling spheres. However, these calculations are valid only when the lubrication gap thickness is much larger than $O(Kn)$ and thus do not accurately capture the non-continuum hydrodynamics. \cite{sundar96non} derived the non-continuum lubrication force by solving the linearized Boltzmann equation for the non-continuum flow in a channel (a local geometry of the lubrication gap) when the gap between the surfaces is less than or equal to $O(Kn)$. Utilizing the work of Sundararajakumar \& Koch, recently, \citet{dhanasekaran2021collision} calculated the modification of axisymmetric mobilities due to non-continuum lubrication interactions.

Important length scales involved in the problem are - radius of the larger particle $a_1$ and the Kolmogorov length scale $\eta = (\nu_f^3/\epsilon)^{1/4}$ (i.e., the length scale of the smallest turbulent eddy). Collisions between particles primarily take place at the smallest scales of turbulence as the shear rates are highest at the smallest scales (see \citealt{tennekes1972first}). The Kolmogorov length scales $\eta$ for turbulence in many industrial applications and clouds are of orders $100$ \textmu m and $1$ mm, respectively. Here, we consider sub-Kolmogorov particles with $a_1/\eta \sim 10^{-2}-10^{-3}$. At the same time, we assume that particles are large enough for Brownian diffusion to be negligible at the smallest turbulence scale.

The relevant time scales present in the problem are - the particle viscous relaxation time $\tau_p = 2 \rho_p a_1^2/9 \mu_f$, the Kolmogorov time scale $\tau_{\eta} = (\nu_f/\epsilon)^{1/2} = \Gamma_{\eta}^{-1}$, and the time a particle takes to settle across a Kolmogorov eddy $\tau_v = \eta/U_s$, where $U_s = g\tau_p$ is the Stokes settling speed. Using the first three time scales we can construct two important non-dimensional numbers - the Stokes number $St = \tau_p/\tau_{\eta} = \varGamma_{\eta}\tau_p$ (measures the particle inertia) and the non-dimensional terminal velocity called the settling parameter $Sv = \tau_{\eta}/\tau_v = U_s/u_{\eta} = g\tau_p/u_{\eta}$, where $u_{\eta} = \eta/\tau_{\eta} = (\nu_f \epsilon)^{1/4}$ is the Kolmogorov velocity. At sub-Kolmogorov scales, fluid inertia is negligible, and the influence of particle inertia on the collision dynamics is insignificant for smaller particles (radii less than 25 \textmu m) and particle pairs with modest size differences as discussed in \cite{dhanasekaran2021turbulent}. In fact, small $St$ is typical for cloud droplets at the lower end of the `size gap' regime and aerosols in industrial reactors (see \citealt{ayala2008effects}). The direct numerical simulation performed by \cite{ireland2016aeffect} suggests that the effect of particle inertia on the particle relative velocity is negligible for $St < 0.2$, and thus we can use the mobility formulation like $St=0$ problems. However, particle inertia will play an important role in collision rates for droplets with larger radii. The clustering of inertial particles in the low vorticity regions leads to enhancement of pair distribution function and hence results in increased collision rate (see DNS study of \citealt{sundaram1997collision,reade2000effect,ireland2016aeffect,dhariwal2018small} and the theoretical analysis of \citealt{zaichik2003pair,chun2005clustering,zaichik2007refinement}).

We consider the rapid-settling regime, $Sv \gg 1$, in which a particle pair traverses a Kolmogorov eddy over a time scale much shorter than the Kolmogorov time scale. Consequently, the turbulent velocity gradients experienced by the particles decorrelate rapidly along the settling trajectory, a key assumption underlying the formulation developed in this study. In a comprehensive review of particle--turbulence interactions, \citet{vaillancourt2000review} compared the region of the ($St, Sv$) parameter space relevant to cloud droplets with that explored in previous laboratory experiments and direct numerical simulations, highlighting that many engineering studies considered parameter regimes substantially different from those encountered in atmospheric clouds. Likewise, \citet{devenish2012droplet} summarized the ranges of $St$ and $Sv$ for individual cloud droplets over representative droplet sizes and turbulent dissipation rates in warm clouds, demonstrating that cloud droplets typically satisfy $St < 1$, whereas $Sv$ may be of order unity or significantly larger depending on the droplet size and turbulence intensity. The Froude number, $Fr = St/Sv = \varGamma_{\eta}^2\eta/g$, provides an alternative measure of the relative importance of turbulence and gravity that is independent of the size and properties of the particles. Using direct numerical simulations, \citet{ireland2016beffect} showed that gravitational effects on inertial clustering are negligible for particles with $St < 0.3$ when $Fr = 0.052$. More recently, \citet{rani2019clusteringa,rani2019clusteringb} demonstrated that, in the rapid-settling limit, gravitational settling can substantially modify the sampling of turbulent velocity gradients even for weakly inertial particles ($St < 0.3$). They developed a theory to quantify the role of gravitational settling on inertial clustering in the rapid-settling limit and compared it with their DNS study for $\epsilon = 10^{-2}$ m$^2$s$^{-3}$, corresponding to $Fr = 0.052$. They reported a significant change in the clustering exponent relative to the case without gravity. 

The characterization of the background turbulence is essential for the accurate prediction of collision rates. The turbulence in clouds is characterized by large Taylor microscale Reynolds number ($Re_{\lambda} \sim 10^4$), low dissipation rates (compared to several common engineering flows), moderate velocity fluctuations and strong intermittency (see \citealt{shaw2003particle}). The dissipation rate $\epsilon$ depends on the cloud type and age; for example, $\epsilon \sim 10^{-1}$ m$^2$s$^{-3}$ in cumulonimbus clouds, $\epsilon \sim 10^{-2}$ m$^2$s$^{-3}$ in cumuli and $\epsilon \sim 10^{-3}$ m$^2$s$^{-3}$ in stratocumuli (see Grabowski \& Wang 2013). The r.m.s. velocity $u'$ is typically around $1$  ms$^{-1}$ and tends to increase with $\epsilon$. DNS of the background turbulence is too expensive, and also, to access cloud-relevant $Re_{\lambda}$ one would be testing the limits of present computational capabilities. In the current work, we require only the energy spectrum of the background turbulence. We will use a model energy spectrum from \cite{pope_2000} for homogeneous isotropic turbulence that agrees well with the properties of small-scale turbulence in high Reynolds number flows.

Since the particle radii are much smaller than the Kolmogorov length scale, the fluid velocity varies approximately linearly over the scale of a particle pair. Consequently, the particle pair experiences a stochastically varying linear flow whose velocity gradient tensor, $\boldsymbol{\varGamma}(t)$, evolves according to the underlying turbulent fluctuations. In addition, the particle Reynolds number based on the terminal settling velocity satisfies $Re_p = a_1 U_s/\nu_f \ll 1$, ensuring that the Stokes equations govern the flow field around each particle. Owing to the linearity of the Stokes equations, the relative velocity of two hydrodynamically interacting inertialess particles settling in turbulence can be expressed as the superposition of the relative velocities induced by the local turbulent shear and differential gravitational settling. The turbulent-shear-induced relative velocity depends on the instantaneous velocity-gradient tensor $\boldsymbol{\varGamma}(t)$, and therefore the evolution of the particle separation requires a statistical description of the velocity gradient sampled along the particle trajectory.

In the limit $Sv \ll 1$, the settling velocity is small compared with the Kolmogorov velocity, so the particle pair closely follows the fluid motion. Consequently, the velocity gradient sampled by the particles can be approximated by the fluid Lagrangian velocity gradient. Many previous studies have developed the model for the stochastic velocity gradient along a fluid trajectory (see \citealt{brunk1998turbulent,girimaji1990diffusion,pereira2018multifractal}). In particular, \citet{dhanasekaran2021turbulent} employed the stochastic velocity-gradient model of \citet{girimaji1990diffusion}, in which the non-Gaussian nature of turbulence is represented through a log-normal model for the pseudo-dissipation rate, to compute turbulent collision rates using a trajectory analysis. In contrast, the present work considers the rapid-settling regime ($Sv \gg 1$), in which the particle pair traverses a Kolmogorov eddy on a time scale much shorter than the eddy turnover time. In this limit, the velocity-gradient tensor must be sampled along the settling trajectory rather than along a fluid trajectory. As shown subsequently, the corresponding velocity-gradient autocorrelation function can be directly linked to the energy spectrum of homogeneous isotropic turbulence, providing a basis for deriving the hydrodynamic diffusivity and the relative drift velocity governing the pair probability distribution.

The strain experienced by a pair settling rapidly through a turbulent eddy is small. Therefore, the relative separation of a pair caused by turbulent shear alone does not change significantly over a correlation time of the background flow and a diffusive process can characterize the relative particle motion caused by many uncorrelated fluctuations. The two-time correlation of the relative velocity due to turbulent shear, integrated over time, measures the diffusivity. Hydrodynamic interactions make the relative velocity due to turbulent shear nonsolenoidal, resulting in a net drift between the pair. Therefore, we can write the relative particle motion in terms of a pair probability conservation equation containing a diffusivity tensor and combined relative velocity due to drift and differential sedimentation. The pair probability $P(\textit{\textbf{r}},\textit{t})$ describes the probability of finding a particle pair separated by the vector $\textit{\textbf{r}}$ at time $\textit{t}$. We will solve the pair probability equation using a semi-analytical method, exploiting the axisymmetry of the problem, to calculate the collision rate. 

The steady-state pair probability and the collision rate in the absence of hydrodynamic interactions depends on the relative strength of differential sedimentation and turbulence $Q = (4 \rho_p g (a_1^2 - a_2^2)/[9 \mu_f])/(\varGamma_{\eta}(a_1+a_2))$, the settling parameter $Sv$, and the Taylor microscale Reynolds number $Re_{\lambda}$. The expression for $Q$ can also be written as $Q=2(1-\kappa)Sv(\eta/a_1)$, where $\kappa = a_2/a_1$ is the size ratio. This form of $Q$ indicates that having $Q$ of $O(1)$ when $Sv \gg 1$ requires a nearly equal-sized particle pair. Interestingly, condensation leads to nearly monodisperse drops, and thus coalescence among nearly equal size drops with small $St$ are common in the lower range of the `size gap' \citep{dhanasekaran2025effect}. Moreover, $\epsilon$ in clouds is small so that $Fr$ is small, and $Sv$ can be large even for drop sizes that are not very large. This motivates us to develop a theory that predicts the collision dynamics of spheres when $Sv \gg 1$, $St \ll 1$, and $Q=O(1)$.

We calculate the pair probability and the ideal collision rate for different values of a ``turbulent'' P\'eclet number ($\mathcal{P}e$), characterizing the competition between the drift and diffusive fluxes. In \S \ref{Problem formulation}, we will derive the explicit expression of $\mathcal{P}e$ in terms of $Q$, $Sv$, and $Re_{\lambda}$. Hydrodynamic interactions impede the collision process. The collision efficiency, defined as the ratio of collision rate with interactions to the ideal collision rate, is used to quantify the effects of interparticle interactions. In this case, the pair probability and the collision efficiency additionally depend on the size ratio $\kappa$ and the Knudsen number $Kn$. We present results for selected values of $\kappa$ and $Kn$ to provide qualitative insights.

The integral of the radial flux of the pair probability over the collision surface gives the collision rate between a pair of spheres. We assume the probability of finding a pair at the collision radius is zero, and thus the relative velocity does not contribute to the radial flux. The diffusive flux is the sole contributor to the calculation of collision rate.

In the following sections, we will investigate the collision of rapidly settling particles in a turbulent flow mediated by non-continuum hydrodynamic interactions. In \S \ref{Problem formulation}, we will derive a Fokker-Planck equation for the pair probability, the pair diffusivity, and the relative drift velocity. We will present a semi-analytical method for solving the pair probability equation in \S \ref{Semi-analytical solution}. Results for the ideal collision rate with no interparticle interactions will be presented in \S \ref{Ideal collision rate}. In \S \ref{Collision with hydrodynamic interactions}, we will present the collision efficiency for different combinations of size ratio and the Knudsen number. Finally, in \S \ref{Summary and conclusions}, we will summarize our results and discuss their implications.

\section{Problem formulation}\label{Problem formulation}
\subsection{Derivation of the pair probability conservation equation}\label{Pair-probability derivation}
The relative velocity $\textbf{\textit{v}}$ between two inertialess spheres sedimenting in a turbulent flow can be expressed as
\begin{equation}
    \textbf{\textit{v}}(t) = \textbf{\textit{w}}(t) + \textbf{\textit{V}}^g,
\label{Relative_velocity_between_two_particles}
\end{equation}
where $\textbf{\textit{w}}(t)$ and $\boldsymbol{V}^{g}$ denote the contributions to the relative velocity from turbulent shear and gravity, respectively. Since sub-Kolmogorov particle pairs experience turbulence as a fluctuating local linear field, $\textbf{\textit{w}}(t)$ can be written as
\begin{equation}
   w_i(t) = \varGamma_{il}(t)r_l-\Big[A\frac{r_ir_k}{r^2}+B\left(\delta_{ik}-\frac{r_ir_k}{r^2}\right)\Big]S_{kl}(t)r_l,
\label{Relative_velocity_due_to_turbulent_shear} 
\end{equation}
where $\varGamma_{il}(t)$ is the statistically stationary fluctuating velocity gradient arising from the background turbulence, $S_{kl}(t)=\left(\varGamma_{kl}(t)+\varGamma_{lk}(t)\right)/2$ is the instantaneous rate of strain, $\delta_{ik}$ is the rank-2 identity tensor, $A$ and $B$ are respectively the axisymmetric and asymmetric mobility functions for two hydrodynamically interacting spherical particles in a linear flow field (see \citealt{batchelor1972hydrodynamic}). The characteristic velocity gradient and rate of strain are defined as: $\varGamma = ( \left<\varGamma_{ij}\varGamma_{ij}\right>)^{1/2}$ and $S = (\left<S_{ij}S_{ij}\right>)^{1/2}$, where the $\left<\cdot\right>$ denotes ensemble averaging over many independent realisations of the background fluctuations. Similarly, we denote the characteristic rotation rate by $R$. The relative velocity due to differential sedimentation, $\textbf{\textit{V}}^g$, is deterministic, and from \citet{batchelor1982sedimentation1} we have
\begin{equation}
    V_i^g = \frac{2\rho_p g_k\left(a_1^2-a_2^2\right)}{9\mu_f}\Big[L\frac{r_ir_k}{r^2}+M\left(\delta_{ik}-\frac{r_ir_k}{r^2}\right)\Big],
\label{Relative_velocity_due_to_differential_sedimentation}
\end{equation}
where $L$ and $M$ are respectively the axisymmetric and asymmetric mobility functions associated with the differential settling of a particle pair under gravity in a quiescent fluid. The radial and polar components of $\textbf{\textit{V}}^g$, which will be used in the subsequent analysis, are given by
\begin{align}
   V_r^g &= -\frac{2\rho_p g\left(a_1^2-a_2^2\right)}{9\mu_f} L\cos\theta,
\label{Radial_relative_velocity_due_to_differential_sedimentation}\\
V_{\theta}^g &= \frac{2\rho_p g\left(a_1^2-a_2^2\right)}{9\mu_f} M\sin\theta,
\label{Polar_relative_velocity_due_to_differential_sedimentation}
\end{align}
where $\theta$ is the angle between $\boldsymbol{r}$ (i.e., the radial vector giving the position of the centre of the satellite sphere with respect to the center of the primary sphere located at the origin) and the vertical axis ($-\boldsymbol{g}$ direction).

These  mobility functions ($A$, $B$, $L$, $M$) depend on the particle size ratio $\kappa = a_2/a_1$ and the dimensionless centre-to-centre distance $r/a^*$. The methods for computing these mobilities together with their asymptotic forms in the limits of small and large separation distances are available in  \citet{batchelor1972hydrodynamic,batchelor1982sedimentation1,jeffrey1984calculation,zhang1991rate,jeffrey1992calculation} and \citet{wang1994collision}. In this analysis, we employ the uniformly valid solutions for $A$ and $L$ developed by \citet{dhanasekaran2021collision}, which consider continuum lubrication interactions for $s > \textrm{O}(Kn)$ and non-continuum lubrication interactions for $s \leq \textrm{O}(Kn)$, where $s = (r-(a_1+a_2))/a^* = (r/a^*)-2$ is the dimensionless surface-to-surface distance between the spheres. Since continuum asymmetric mobilities ($B$ and $M$) approach finite values for $s \rightarrow 0$, we expect that continuum breakdown will not strongly influence the asymmetric relative motions of an inertialess particle pair. Accordingly, we consider continuum hydrodynamics for asymmetric mobilities at all separation distances. However, for inertial particles, non-continuum lubrication forces in the tangential direction may become important for accurate collision rate calculations \citep{how2021non}.

The particle relative velocity given by \eqref{Relative_velocity_between_two_particles} can therefore be expressed explicitly as
\begin{align}
    {v}_i(t) = \dot{r}_i &=\varGamma_{il}(t)r_l-\Big[A\frac{r_ir_k}{r^2}+B\left(\delta_{ik}-\frac{r_ir_k}{r^2}\right)\Big]S_{kl}(t)r_l \nonumber \\  
&+\frac{2\rho_p g_k\left(a_1^2-a_2^2\right)}{9\mu_f}\Big[L\frac{r_ir_k}{r^2}+M\left(\delta_{ik}-\frac{r_ir_k}{r^2}\right)\Big].
\label{Relative_velocity}
\end{align}

The relative velocity (\ref{Relative_velocity}) shows that the particle pair's trajectory is governed by the deterministic differential sedimentation $\textbf{\textit{V}}^g$ and the stochastic turbulent-shear-induced velocity $\textbf{\textit{w}}(t)$, which depends on the fluctuating velocity gradient $\boldsymbol{\varGamma}(t)$ sampled along the settling trajectory of the primary particle. In the rapid settling limit ($Sv \gg 1$), the centre of mass of the particle pair traverses a Kolmogorov eddy in a time $\tau_v = \eta/U_s$ that is much shorter than the eddy turnover time $\tau_{\eta}$. As a result, the velocity gradient $\boldsymbol{\varGamma}(t)$ experienced by the pair decorrelates over the short time $\tau_v$, not the Kolmogorov time $\tau_{\eta}$. The total strain accumulated during one correlation time is therefore $\varGamma_{\eta}\tau_v = \varGamma_{\eta}\eta/U_s = u_{\eta}/U_s = Sv^{-1} \ll 1$, confirming that the pair experiences only a small strain before the velocity gradient fluctuations become uncorrelated. 
 
This situation is analogous to the classical turbulent diffusion problem studied rigorously by \cite{kesten1979limit}. In their framework, a particle at position $\boldsymbol{x}$ moves with velocity $\boldsymbol{V}_0 + \varepsilon \boldsymbol{F}(\boldsymbol{x})$, where $\boldsymbol{V}_0$ is a large constant mean velocity, $\boldsymbol{F}(\boldsymbol{x})$ is a zero-mean stationary random field, and $\varepsilon \ll 1$ measures the fluctuation amplitude. As $\varepsilon \rightarrow 0$, the particle's displacement relative to the mean motion converges to a diffusion process with coefficients given by integrals of the velocity correlation function $\left<F_i(\boldsymbol{x})F_j(\boldsymbol{x}+\boldsymbol{V}_0 t)\right>$ evaluated along the mean trajectory. In the present problem, the role of the mean velocity $\boldsymbol{V}_0$ is played by the gravitational settling velocity $\boldsymbol{g}\tau_p$ of the centre of mass, which carries the pair through the turbulent eddies. The turbulent-shear-induced relative velocity $\textbf{\textit{w}}(t)$ plays the role of the fluctuation $\varepsilon \boldsymbol{F}$, with the small parameter $\varepsilon = Sv^{-1}$ representing the total strain. Since many uncorrelated velocity gradient fluctuations contribute to the relative displacement before a collision event occurs, the relative particle motion due to turbulent shear is diffusive. An analogous mechanism underlies the mechanical dispersion of a passive tracer advected through a random porous medium, where the mean flow carries the tracer past many randomly positioned grains, giving rise to an effective diffusivity expressed as an integral of the velocity disturbance correlation function along the mean trajectory (\citealt{koch1985dispersion,koch1987non}).
 
We now derive the drift-diffusion equation governing the pair probability. With respect to the coordinate system attached at the centre of the primary particle, the pair density function $\varOmega(\textit{\textbf{r}},t)$ gives the number density of particle pairs separated by $\textit{\textbf{r}}$ at time $t$ for a given realisation of the turbulent flow. The pair density satisfies the conservation equation:
\begin{equation}
\frac{\partial \varOmega}{\partial t} + \frac{\partial }{\partial r_i}(v_i \varOmega) = 0.
\label{Total_probability_conservation_equation}
\end{equation}
The pair density $\varOmega$ is influenced both by the deterministic differential sedimentation and by the fluctuations of $w_i$, which arise from the turbulent velocity gradient. We define the pair probability $P(\textit{\textbf{r}},t)$ as the ensemble average of $\varOmega(\textit{\textbf{r}},t)$ over realisations of the background turbulent flow:
\begin{eqnarray}
 P(\textit{\textbf{r}},t) = \left<\varOmega(\textit{\textbf{r}},t)\right>.
 \label{Averaged_pair_probability}
\end{eqnarray}
By definition, $P\rightarrow n_1 n_2$ as $|\textit{\textbf{r}}|\rightarrow\infty$. As shown rigorously by \cite{kesten1979limit} for a particle moving through a spatially random field with a large mean velocity, the rapid traversal of many independent correlation lengths renders the ensemble average equivalent to an average over the many independent regions of the random field encountered along the trajectory. Here, the rapidly settling pair samples many independent Kolmogorov eddies, so the ensemble average over turbulent flow realisations is equivalent to averaging over the eddies encountered along the settling path. The fluctuating component of $\varOmega$ is $p = \varOmega - P$.
 
Ensemble-averaging (\ref{Total_probability_conservation_equation}) yields
\begin{eqnarray}
\frac{\partial P}{\partial t} + \frac{\partial }{\partial r_i}(\left<w_i \varOmega\right>) + \frac{\partial }{\partial r_i}(V_i^g P) = 0,
 \label{Ensemble_average_of_total_pair_probability_step1}
\end{eqnarray}
where we have used the fact that the differential sedimentation velocity $V_i^g$ is deterministic. The turbulent-shear-induced relative velocity $w_i$ has zero ensemble mean due to the isotropy of the background turbulence, i.e.\ $\left<w_i\right>=0$. Substituting $\varOmega=P+p$ into (\ref{Ensemble_average_of_total_pair_probability_step1}) and using $\left<w_i P\right> = \left<w_i\right> P = 0$, we obtain
\begin{eqnarray}
\frac{\partial P}{\partial t} + \frac{\partial }{\partial r_i}(\left<w_i p\right>) + \frac{\partial }{\partial r_i}(V_i^g P) = 0.
\label{Reduced_ensemble_average_equation}
\end{eqnarray}
Equation (\ref{Reduced_ensemble_average_equation}) shows that the evolution of $P$ is driven by two mechanisms: the advection due to the deterministic differential sedimentation $V_i^g$, and the correlation $\left<w_i p\right>$ between the turbulent-shear-induced velocity and the fluctuation in the pair density. The latter term will give rise to diffusive and drift contributions, as we now demonstrate.
 
To obtain an expression for $p$, we subtract (\ref{Reduced_ensemble_average_equation}) from (\ref{Total_probability_conservation_equation}), giving
\begin{eqnarray}
\frac{\partial p}{\partial t} + \frac{\partial }{\partial r_i}(w_i P) + \frac{\partial }{\partial r_i}(w_i p - \left<w_i p\right>) + \frac{\partial }{\partial r_i}(V_i^g p) = 0. 
\label{Fluctuation_equation}
\end{eqnarray}
We now use the smallness of the total strain $\varGamma_{\eta}\tau_v = Sv^{-1} \ll 1$ to simplify (\ref{Fluctuation_equation}). The fluctuation $p$ is generated by the $w_i P$ term, which involves the velocity gradient with characteristic magnitude $\varGamma_{\eta}$. Since $p$ is produced over the correlation time $\tau_v$ and must satisfy $\left<p\right>=0$, we have $|p| \sim \varGamma_{\eta}\tau_v |P| = Sv^{-1}|P| \ll |P|$. Consequently, the nonlinear term $w_i p - \left<w_i p\right>$ in (\ref{Fluctuation_equation}) is $O(Sv^{-1})$ smaller than $w_i P$ and may be neglected at leading order.
 
Furthermore, since $p$ varies on the fast time scale $\tau_v$ while $P$ evolves on the much slower collision time scale $\tau_Q \gg \tau_v$, we can treat $P$ as quasi-steady on the time scale of the fluctuations. This separation of time scales -- the essence of the rapid settling limit -- allows us to solve for $p$ by integrating (\ref{Fluctuation_equation}) over the history of the fluctuations:
\begin{eqnarray}
p(\textit{\textbf{r}},t) \approx -\int_{-\infty}^t \frac{\partial }{\partial r_i}\Big(w_i(t') P(\textit{\textbf{r}},t)\Big) dt'.
\label{Explicit_expression_of_fluctuating_pair_probability}
\end{eqnarray}
In writing (\ref{Explicit_expression_of_fluctuating_pair_probability}), we have also neglected the $V_i^g p$ term compared with $\partial p/\partial t$, because the advection due to differential sedimentation acts on the slow collision time scale $\tau_Q$, while $\partial p/\partial t \sim p/\tau_v$ reflects the rapid decorrelation. Physically, equation (\ref{Explicit_expression_of_fluctuating_pair_probability}) states that the fluctuation $p$ at time $t$ is the cumulative result of the turbulent-shear-induced flux $w_i P$ acting over all previous times.
 
Substituting (\ref{Explicit_expression_of_fluctuating_pair_probability}) into the correlation $\left<w_i p\right>$ appearing in (\ref{Reduced_ensemble_average_equation}) and separating the resulting expression into diffusive and drift contributions, the pair probability equation becomes:
\begin{equation}
\frac{\partial P}{\partial t} + \frac{\partial}{\partial r_i}\left[\left(V_i^g + V_i^H \right)P - D_{ij}^H \frac{\partial P}{\partial r_j}\right] = 0.
\label{Final_pair_probability_equation}
\end{equation}
The above equation governs the relative motion between two particles settling in homogeneous isotropic turbulence in the limit of small total strain ($Sv^{-1} \ll 1$). This is the central result of the Kesten--Papanicolaou theory applied to the pair problem: when the settling velocity is large, the stochastic relative motion due to turbulent shear is characterized by a pair diffusivity $D_{ij}^H$ and a relative drift velocity $V_i^H$. These are given by the two-time correlation functions involving the turbulent-shear-induced relative velocity and its divergence, integrated along the settling trajectory:
\begin{align}
    D_{ij}^H&=\int_{-\infty}^t\left<w_i(\boldsymbol{r}(t),\boldsymbol{x}(t),t) w_j(\boldsymbol{r}(t'),\boldsymbol{x}(t'),t')\right>dt',
    \label{Relative_diffusion}\\
    V_i^H&=-\int_{-\infty}^t\left<w_i(\boldsymbol{r}(t),\boldsymbol{x}(t),t)\frac{\partial w_l}{\partial r_l}(\boldsymbol{r}(t'),\boldsymbol{x}(t'),t')\right>dt'.
\label{Relative_drift}
\end{align}
Here $\boldsymbol{x}(t)$ is the position of the primary particle at time $t$, and the dependence of $w_i$ on $\boldsymbol{x}$ reflects the fact that the velocity gradient is sampled along the settling trajectory of the centre of mass. The pair diffusivity (\ref{Relative_diffusion}) is the time integral of the two-time relative velocity correlation function -- the direct analogue of the Green--Kubo relation for diffusion coefficients (see \citealt{koch1987non}). The drift velocity (\ref{Relative_drift}) arises from temporal correlations between the turbulent-shear-induced relative velocity and its divergence; it is non-zero only in the presence of hydrodynamic interactions, which render the relative velocity field $w_i$ non-solenoidal, and it vanishes when $A=B=0$. This mechanism is analogous to the drift in orientation space that accompanies the hydrodynamic rotary diffusion of axisymmetric particles translating through a fixed bed, where the disturbance angular velocity is non-solenoidal \citep{shaqfeh1988effect}. In several previous studies, similar drift-diffusion equations have been derived. \cite{brunk1997hydrodynamic} derived the expressions for the hydrodynamic diffusivity and relative drift velocity for inertialess particle pairs in an isotropic random velocity field, assuming a small total strain $\varGamma_{\eta}\tau_S \ll 1$ in a fluid Lagrangian frame without gravitational settling. \cite{chun2005clustering} and \cite{rani2019clusteringa,rani2019clusteringb} obtained similar drift-diffusion fluxes in the context of inertial particle clustering in homogeneous isotropic turbulence in the absence of hydrodynamic interactions, where the finite particle inertia is responsible for the relative drift. In the present study, as in \citet{rani2019clusteringa}, the small total strain is a direct consequence of the rapid settling ($Sv \gg 1$), which causes the velocity gradient to decorrelate over the short advective time $\tau_v = \eta/U_s \ll \tau_{\eta}$.

\subsection{Hydrodynamic pair diffusivity}\label{pair diffusion calculation}

The relevant term for calculating pair diffusivity is the two-time relative velocity correlation, $\mathcal{H}_{ij} = \left<w_i(\boldsymbol{r}(t),\boldsymbol{x}(t),t) w_j(\boldsymbol{r}(t'),\boldsymbol{x}(t'),t')\right>$. Substituting for the expression of $w_i$ from equation (\ref{Relative_velocity_due_to_turbulent_shear}) we have
\begin{align}
\mathcal{H}_{ij} &= \left<\left[\varGamma_{il}(t)r_l-\left\{A\frac{r_ir_k}{r^2}+B\left(\delta_{ik}-\frac{r_ir_k}{r^2}\right)\right\}S_{kl}(t)r_l\right]\times\right. \nonumber\\
&\left.\left[\varGamma_{jm}(t')r_m-\left\{A\frac{r_jr_m}{r^2}+B\left(\delta_{jm}-\frac{r_jr_m}{r^2}\right)\right\}S_{mp}(t')r_p\right]\right>.
\label{Original_form_of_R_ij}
\end{align}
The change in the relative separation between the particle pair over the velocity gradient correlation time is not significant because they experience small strain while settling rapidly through Kolmogorov eddies. Thus the relative position vectors present in (\ref{Original_form_of_R_ij}) can be assumed to be independent of the instantaneous velocity gradient $\boldsymbol{\varGamma}$ and taken out from the average:
 \begin{align}
 \mathcal{H}_{ij} &= r_lr_m\left<\varGamma_{il}(t)\varGamma_{jm}(t')\right>-r_lr_p\left\{A\frac{r_jr_m}{r^2}+B\left(\delta_{jm}-\frac{r_jr_m}{r^2}\right)\right\}\left<\varGamma_{il}(t)S_{mp}(t')\right> - \nonumber\\
 & r_lr_m\left\{A\frac{r_ir_k}{r^2}+B\left(\delta_{ik}-\frac{r_ir_k}{r^2}\right)\right\}\left<\varGamma_{jm}(t')S_{kl}(t)\right> + \nonumber\\ 
  & r_lr_p\left\{A\frac{r_ir_k}{r^2}+B\left(\delta_{ik}-\frac{r_ir_k}{r^2}\right)\right\} \times \nonumber\\
 & \left\{A\frac{r_jr_m}{r^2}+B\left(\delta_{jm}-\frac{r_jr_m}{r^2}\right)\right\}\left<S_{kl}(t)S_{mp}(t')\right>, 
\label{operation_1_for_R_ij}
 \end{align}
 where $\varGamma_{ij}(t) \equiv \varGamma_{ij}(\boldsymbol{x}(t),t)$ , $\varGamma_{ij}(t') \equiv \varGamma_{ij}(\boldsymbol{x}(t'),t')$ and similar relations are assumed for $S_{ij}$. Now, substituting the expression of $\mathcal{H}_{ij}$ into (\ref{Relative_diffusion}) we have:
\begin{align}
D_{ij}^H &= r_lr_m\Pi_{iljm}-\frac{r_lr_p}{2}\left\{A\frac{r_jr_m}{r^2}+B\left(\delta_{jm}-\frac{r_jr_m}{r^2}\right)\right\}(\Pi_{ilmp}+\Pi_{ilpm})-\nonumber\\
& \frac{r_lr_m}{2}\left\{A\frac{r_ir_k}{r^2}+B\left(\delta_{ik}-\frac{r_ir_k}{r^2}\right)\right\}(\Pi_{jmkl}+\Pi_{jmlk})+\frac{r_lr_p}{4}\left\{A\frac{r_ir_k}{r^2}\right. + \nonumber\\
& \left.B\left(\delta_{ik}-\frac{r_ir_k}{r^2}\right)\right\}\left\{A\frac{r_jr_m}{r^2}+B\left(\delta_{jm}-\frac{r_jr_m}{r^2}\right)\right\} \times  \nonumber\\
& \left(\Pi_{klmp}+\Pi_{klpm}+\Pi_{lkmp}+\Pi_{lkpm}\right).
\label{Detail_expression_of_Diffusivity}
\end{align}
Here $\pmb{\Pi}$ is a rank-4 tensor that depends on the velocity gradient autocorrelation function,
\begin{equation}
\Pi_{ijkl}=\int_{-\infty}^t\left<\varGamma_{ij}(t)\varGamma_{kl}(t')\right>dt'.
\label{Velocity_gradient_autocorrelation_function_1}
\end{equation}
In the rapid settling regime, particle settling velocity is much greater than the Kolmogorov-scale fluid velocity (i.e., $g\tau_p \gg u_{\eta}$) and therefore we need to calculate this autocorrelation function along a settling trajectory. The simplification for the $\pmb{\Pi}$ tensor due to the rapid settling assumption was recently done by \cite{rani2019clusteringa}, and we will reproduce below the primary elements of the derivation for the sake of completeness. The primary particle's position at time $t'$ can be approximated as $\boldsymbol{x}(t')=\boldsymbol{x}(t)+\boldsymbol{g}\tau_p(t'-t)$ and thus
$\left<\varGamma_{ij}(t)\varGamma_{kl}(t')\right> = \left<\varGamma_{ij}(\boldsymbol{x}(t),t)\varGamma_{kl}(\boldsymbol{x}(t'),t')\right> = \left<\varGamma_{ij}(\boldsymbol{x}(t),t)\varGamma_{kl}(\boldsymbol{x}(t)+\boldsymbol{x}_g,t)\right>,$ 
where $\boldsymbol{x}_g = \boldsymbol{g}\tau_p(t'-t)$. Now, we can express $\left<\varGamma_{ij}(\boldsymbol{x}(t),t)\varGamma_{kl}(\boldsymbol{x}(t)+\boldsymbol{x}_g,t)\right>$ as
\begin{equation}
\left<\varGamma_{ij}(\boldsymbol{x}(t),t)\varGamma_{kl}(\boldsymbol{x}(t)+\boldsymbol{x}_g,t)\right> =\left< \frac{\partial u_i}{\partial x_j}(\boldsymbol{x}) \frac{\partial u_k}{\partial x_l}(\boldsymbol{x} + \boldsymbol{x}_g)\right>.
\label{Velocity_gradient_autocorrelation_function_2}
\end{equation}
Expressing the fluid velocities $u_i$ and $u_k$ in terms of Fourier series we have
\begin{align}
\frac{\partial u_i}{\partial x_j}(\boldsymbol{x}) &= \int \textsf{i}  k_j \hat {u}_i(\textbf{\textit{k}})e^{\textsf{i}\textbf{\textit{k}}\boldsymbol{\cdot}\boldsymbol{x}}d\textbf{\textit{k}},
\label{Fourier_mode_velocity_1}\\
\frac{\partial u_k}{\partial x_l}(\boldsymbol{x}+\boldsymbol{x}_g) &= \int \textsf{i}  k_l' \hat{u}_k(\textbf{\textit{k}}')e^{\textsf{i}\textbf{\textit{k}}'\boldsymbol{\cdot}(\boldsymbol{x} + \boldsymbol{x}_g)}d\textbf{\textit{k}}',
\label{Fourier_mode_velocity_2}
\end{align}
where $\textsf{i}=\sqrt{-1}$, $\textbf{\textit{k}}$ and $\textbf{\textit{k}}'$ are wavenumber vectors, and $\hat {u}_i(\textbf{\textit{k}})$ is a complex Fourier coefficient of the fluid velocity $u_i$ at wavenumber \textbf{\textit{k}}.
Using the homogeneity, we average the above correlation over $\boldsymbol{x}$-space:
\begin{align}
\left< \left< \frac{\partial u_i}{\partial x_j}(\boldsymbol{x}) \frac{\partial u_k}{\partial x_l}(\boldsymbol{x} + \boldsymbol{x}_g)\right>_\mathcal{L}\right>  &= -\int\int d\textbf{\textit{k}} d\textbf{\textit{k}}' k_j k_l' \left< \hat{u}_i(\textbf{\textit{k}}) \hat{u}_k(\textbf{\textit{k}}')\right> \left< e^{\textsf{i}\textbf{\textit{k}}\boldsymbol{\cdot}\boldsymbol{x}} e^{\textsf{i}\textbf{\textit{k}}'\boldsymbol{\cdot}(\boldsymbol{x} + \boldsymbol{x}_g)} \right>_\mathcal{L} \nonumber\\
&=-\int\int d\textbf{\textit{k}} d\textbf{\textit{k}}' k_j k_l' \left< \hat{u}_i(\textbf{\textit{k}}) \hat{u}_k(\textbf{\textit{k}}')\right>  \delta(\textbf{\textit{k}} + \textbf{\textit{k}}')e^{\textsf{i}\textbf{\textit{k}}'\boldsymbol{\cdot}\boldsymbol{x}_g} \nonumber\\
&= \int d\textbf{\textit{k}}  k_j k_l \left< \hat{u}_i(\textbf{\textit{k}}) \hat{u}_k(-\textbf{\textit{k}})\right>  e^{-\textsf{i}\textbf{\textit{k}}\boldsymbol{\cdot}\boldsymbol{x}_g} \nonumber\\
&= \int d\textbf{\textit{k}} k_j k_l \left< \hat{u}_i(\textbf{\textit{k}}) \hat{u}_k^*(\textbf{\textit{k}})\right> e^{-\textsf{i}\textbf{\textit{k}}\boldsymbol{\cdot}\boldsymbol{x}_g} \nonumber\\
&= \int d\textbf{\textit{k}} k_j k_l \varPhi_{ik}(\textbf{\textit{k}}) e^{-\textsf{i}\textbf{\textit{k}}\boldsymbol{\cdot}\boldsymbol{x}_g} \nonumber\\
&= \int d\textbf{\textit{k}} k_j k_l \varPhi_{ik}(\textbf{\textit{k}}) e^{-\textsf{i}\textbf{\textit{k}}\boldsymbol{\cdot}\boldsymbol{g}\tau_p(t'-t)},
\label{volume_average_of_autocorrelation}
\end{align}
where $\left<\cdot\cdot\cdot\right>_\mathcal{L}$ denotes the volume averaging, which is performed over an infinite volume (i.e., $\mathcal{L}\rightarrow \infty$), $\delta(\cdot\cdot\cdot)$ denotes Dirac delta function, superscript * denotes complex conjugate, and  $\varPhi_{ik}(\textbf{\textit{k}})$ is the velocity-spectrum tensor. The velocity spectrum tensor is related to the turbulence energy-spectrum as given below (see \citealt{pope_2000}): 
\begin{equation}
    \varPhi_{ik}(\textbf{\textit{k}}) = \frac{E(\textit{k})}{4\pi \textit{k}^2} \left(\delta_{ik}-\frac{\textit{k}_i \textit{k}_k}{\textit{k}^2}\right), 
\label{Relation_between_velocity_spectrum_tensor_and_energy_spectrum}
\end{equation}
where $E(\textit{k})$ is the turbulence energy-spectrum. Substituting this expression of $\varPhi_{ik}(\textbf{\textit{k}})$ into (\ref{volume_average_of_autocorrelation}) we have:  
\begin{equation}
    \left<\varGamma_{ij}(t)\varGamma_{kl}(t')\right> = \int d\textbf{\textit{k}} \frac{E(\textit{k})}{4\pi \textit{k}^2} \left(\delta_{ik}-\frac{\textit{k}_i \textit{k}_k}{\textit{k}^2}\right) k_j k_l e^{-\textsf{i}\textbf{\textit{k}}\boldsymbol{\cdot}\boldsymbol{g}\tau_p(t'-t)}.
\label{Final_expression_of_autocorrelation}
\end{equation}
Letting $t=0$ and using (\ref{Final_expression_of_autocorrelation}), the $\pmb{\Pi}$ tensor becomes 
\begin{align}
    \Pi_{ijkl} &= \int d\textbf{\textit{k}} \frac{E(\textit{k})}{4\pi \textit{k}^2} \left(\delta_{ik}-\frac{\textit{k}_i \textit{k}_k}{\textit{k}^2}\right) k_j k_l  \int_{-\infty}^0 e^{-\textsf{i}\textbf{\textit{k}}\boldsymbol{\cdot}\boldsymbol{g}\tau_p t'} dt' \nonumber\\
   &= \int d\textbf{\textit{k}} \frac{E(\textit{k})}{4\pi \textit{k}^2} \left(\delta_{ik}-\frac{\textit{k}_i \textit{k}_k}{\textit{k}^2}\right) k_j k_l \left\{\frac{1}{2} \delta \left[-\frac{\textbf{\textit{k}}\boldsymbol{\cdot}\boldsymbol{g}\tau_p}{2\pi}\right]- \frac{1}{\textsf{i} \textbf{\textit{k}}\boldsymbol{\cdot}\boldsymbol{g}\tau_p}\right\},
\label{Expression_of_PI_tensor_1}
\end{align}
where we have used the Fourier transform identity for $\int_{-\infty}^0 e^{-\textsf{i}\textbf{\textit{k}}.\boldsymbol{g}\tau_p t'} dt'$. It can be proved that the second term in the curly bracket does not contribute to the above integral. So, the above integral will exist when $\textbf{\textit{k}}.\boldsymbol{g}=0$. Clearly, $\textbf{\textit{k}}=\pmb{\xi}=(\xi_1, \xi_2, 0)$ satisfies this property. Using the sifting property and the transformation rule (i.e., $\delta(bx)=(1/|b|)\delta(x)$) of the Dirac delta function, the above integral takes the following form:
\begin{equation}
   \Pi_{ijkl} = \frac{\pi}{g\tau_p}\int d\pmb{\xi}\,\frac{E(\xi)}{4\pi\xi^2}\left(\delta_{ik}-\frac{\xi_i\xi_k}{\xi^2}\right)\xi_j\xi_l.
\label{Expression_of_PI_tensor_2} 
\end{equation}
$\pmb{\Pi}$ is a rank-4 axisymmetric tensor and thus it assumes the following general form -
\begin{align}
\Pi_{ijkl}&=\alpha_1\delta_{ij}\delta_{kl}+\alpha_2\delta_{il}\delta_{jk}+\alpha_3\delta_{ik}\delta_{jl}+\alpha_4\delta_{i3}\delta_{j3}\delta_{k3}\delta_{l3}+\alpha_5\delta_{i3}\delta_{j3}\delta_{kl}+\alpha_6\delta_{i3}\delta_{k3}\delta_{jl} \nonumber\\
& +\alpha_7 \delta_{i3}\delta_{l3}\delta_{jk}+\alpha_8\delta_{j3}\delta_{k3}\delta_{il}+\alpha_9\delta_{j3}\delta_{l3}\delta_{ik}+\alpha_{10}\delta_{k3}\delta_{l3}\delta_{ij},
\label{General_expression_of_PI_tensor} 
\end{align}
where $\alpha_1, \hdots, \alpha_{10}$ are scalar constants. To further simplify $\pmb{\Pi}$, we apply the incompressibility constraint $\Pi_{iikl}=\Pi_{ijkk}=0$, the symmetries of the problem $\Pi_{ijkl}=\Pi_{kjil}=\Pi_{klij}=\Pi_{ilkj}$ and $\Pi_{i3kl}=\Pi_{ijk3}=0$. After simplifying, we get:
\begin{align}
    \Pi_{ijkl}&=\alpha_1\left[\delta_{i3}\delta_{j3}\delta_{k3}\delta_{l3}-\delta_{i3}\delta_{k3}\delta_{jl}\right]+\alpha_2\left[\delta_{il}\delta_{kj}+\delta_{ij}\delta_{lk}-3\delta_{ik}\delta_{lj}-\delta_{i3}\delta_{l3}\delta_{kj}\right.\nonumber\\
 &- \left.\delta_{l3}\delta_{k3}\delta_{ij} - \delta_{i3}\delta_{j3}\delta_{lk}-\delta_{k3}\delta_{j3}\delta_{il}+2\delta_{i3}\delta_{k3}\delta_{lj}+3\delta_{j3}\delta_{l3}\delta_{ik}\right]
\label{Expression_of_PI_tensor_after_appliying_constraints}
\end{align}
Comparing (\ref{Expression_of_PI_tensor_2}) with (\ref{Expression_of_PI_tensor_after_appliying_constraints}), we can express $\alpha_1$ and $\alpha_2$ in terms of the turbulence energy-spectrum as follows:
\begin{equation}
    \frac{\alpha_1}{3}=\alpha_2=-\frac{\pi}{16g\tau_p}\int_0^{\infty}\xi\, d\xi E(\xi).
\label{General_expressions_of_alpha1_and_alpha2}
\end{equation}
It is evident from the above expression that numerical values of $\alpha_1$ and $\alpha_2$ are negative. With an assumed form of the energy-spectrum and a given value of $Re_{\lambda}$, we can show that $\alpha_2$ scaled by $\varGamma_{\eta}$ is inversely proportional to $Sv$ (details are provided in Appendix \ref{appB}). On substituting (\ref{Expression_of_PI_tensor_after_appliying_constraints}) into (\ref{Detail_expression_of_Diffusivity}) and using the constraints on the $\pmb{\Pi}$ tensor we have:
\begin{align}
    \frac{D_{ij}^H}{\alpha_2}&=\delta_{i3}\delta_{j3}(3r_3^2-r^2)+2r_ir_j+3\delta_{ij}(r_3^2-r^2)-2r_3(r_j\delta_{i3}+r_i\delta_{j3})\nonumber\\
 &-\frac{A-B}{r^2}\left\{2r_ir_j+3r_3(r_i\delta_{j3}+r_j\delta_{i3})\right\}(r_3^2-r^2)\nonumber\\
 &-2B\left\{(3r_3^2-r^2)\delta_{i3}\delta_{j3}-r_3(r_i\delta_{j3}+r_j\delta_{i3})+(r_3^2-r^2)\delta_{ij}\right\} \times \nonumber\\
 &\frac{(A-B)^2}{r^4}r_ir_j(r_3^2-r^2)(3r_3^2+r^2)\nonumber\\
 & +\frac{B(A-B)}{r^2}\left\{r_3(3r_3^2-r^2)(r_i\delta_{j3}+r_j\delta_{i3})-2r_ir_j(r^2+r_3^2)\right\}\nonumber\\
 &+B^2(3r_3^2\delta_{i3}\delta_{j3}-r_3(r_i\delta_{j3}+r_j\delta_{i3})-r^2\delta_{ij}),
\label{End_expression_of_Diffusivity}
\end{align}
where $r_3=r\cos\theta$. As we are studying an axisymmetric problem, it is sufficient to carry out our analysis in the $(r,\theta)$ plane. The components of the diffusivity in spherical coordinates obtained from (\ref{End_expression_of_Diffusivity}) are:
\begin{align}
   D_{rr}^H &= -\frac{1}{2}(A-1)^2 \sin ^2\theta \left\{3 \cos 2 \theta +5 \right\} \alpha_2 r^2, \label{Diffusivity_D_rr}\\
  D_{r \theta}^H &= -\frac{1}{4} (A-1) \sin 2 \theta \left\{ 3 (B-1) \cos 2 \theta +B+3 \right\}  \alpha_2 r^2,
\label{Diffusivity_D_rtheta}\\
  D_{\theta \theta}^H &= -\frac{1}{8} \left\{ 3 (B-1)^2 \cos 4 \theta + 16 (B-1) \cos 2 \theta + 5B (B-2) + 13\right \} \alpha_2 r^2. 
\label{Diffusivity_D_thetatheta} 
\end{align}
It is important to note that the pair diffusion coefficients in this problem are functions of both $r$ and $\theta$ and are proportional to $\alpha_2 r^2$. By setting $A=B=0$ in the above equations, we can recover the form of the pair diffusion coefficients derived in \cite{rani2019clusteringa} where they ignored the hydrodynamic interactions between the spherical particle pair.

\subsection{Drift velocity} \label{Drift velocity calculation}

When a diffusivity varies with position, it is common for there to be a drift velocity from regions of higher to lower diffusivity. Hydrodynamic drift in the pair probability equation of a rapidly settling particle pair in a turbulent flow is akin to the phenomenon called thermophoresis found for Brownian particles subjected to a temperature gradient. The flux of particles is higher in the hot region than in the cold region because the Brownian diffusivity is directly proportional to the temperature. Therefore, to have zero flux at the steady-state, the particles must be concentrated in the colder region, suggesting a ``drift'' of particles from the hotter region. In this problem, the hydrodynamic interactions between the particle pairs bring about the drift velocity term in the pair probability equation. The presence of position-dependent diffusivity does not guarantee the presence of drift.  For example, there is no drift for inertialess particle pairs with no hydrodynamic interactions rapidly settling in isotropic turbulence, although the relative diffusivity still depends on position. Mathematically, the correlation of the turbulent-shear-induced relative velocity and its divergence yields relative drift between the particle pair. Let us denote this correlation by $\mathscr{D}_i$ (i.e., $\mathscr{D}_i = \left< w_i(\boldsymbol{r}(t),\boldsymbol{x}(t),t) \frac{\partial w_l}{\partial r_l}(\boldsymbol{r}(t'),\boldsymbol{x}(t'),t')\right>$). The non-interacting particles will follow the fluid trajectories and thus $\partial w_l/\partial r_l=0$ in that case. With hydrodynamic interactions, the divergence term is given by (see \citealt{batchelor1972determination})
\begin{equation}
    \frac{\partial w_l}{\partial r_l} = W \frac{r_mS_{mp}r_p}{r^2},
\label{Divergence_with_interactions}
\end{equation}
where
\begin{equation}
    W = -\left\{3(A-B)+r\frac{dA}{dr}\right\}.
\label{Expression_of_W}
\end{equation}
Using (\ref{Relative_velocity_due_to_turbulent_shear}) and (\ref{Divergence_with_interactions}), $\mathscr{D}_i$ becomes
\begin{equation}
    \mathscr{D}_i = W\frac{r_lr_mr_p}{r^2} \Big[ \left< \varGamma_{il}(t)S_{mp}(t')\right>-\left\{A\frac{r_ir_k}{r^2}+B\left(\delta_{ik}-\frac{r_ir_k}{r^2}\right)\right\} \left<S_{kl}(t)S_{mp}(t') \right> \Big]
\label{Expression_of_M_ij}
\end{equation}
Substituting the above expression into (\ref{Relative_drift}) and using (\ref{Expression_of_PI_tensor_after_appliying_constraints}) we have
\begin{align}
  V_i^H =& -\int_{-\infty}^t \mathscr{D}_i(t,t')dt' \nonumber \\ = & -W\Big[-(3\delta_{i3}r\cos\theta+r_i)\sin^2\theta+r_i(A-B)(3\cos^2\theta+1)\sin^2\theta \nonumber\\
& -B\left\{r\cos\theta\delta_{i3}(3\cos^2\theta-1)-r_i(1+\cos^2\theta)\right\}\Big]\alpha_2.
 \label{Explicit_expression_of_drift_velocity}  
\end{align}
From (\ref{Explicit_expression_of_drift_velocity}), we have the following expressions for the radial and polar components of the relative drift velocity
\begin{align}
    V_r^H &= -W(A-1)\sin^2\theta(3\cos^2\theta+1)\alpha_2 r,
\label{Radial_drift}\\
    V_{\theta}^H &= -\frac{W}{2} \sin2\theta\left\{3\sin^2\theta+B(3\cos^2\theta-1)\right\}\alpha_2 r.
\label{Polar_drift}
\end{align}
The radial component of the relative drift is always inward since $W$ is positive and $(A-1)$, $\alpha_2$ are always negative. Since $A=O(1/r^3)$ and $B=O(1/r^5)$ at large separations, $V_r^H$ has an $O(1/r^2)$ far-field decay. For the continuum lubrication interactions, $A \rightarrow 1$, and $B$ reaches a constant value as the separation goes to zero. Therefore $V_r^H$ goes to zero as the particles approach contact for the continuum case (see \citealt{brunk1997hydrodynamic}). On the other hand, $1-A = O(1/\ln(\ln(Kn/(r-2))))$ in the non-continuum lubrication regime which leads to $V_r^H = O(1/[(r-2)\ln(1/(r-2))\{\ln(\ln(Kn/(r-2))))\}^3])$. So, $V_r^H$ approaches $-\infty$ at contact for the non-continuum lubrication interactions. The drift velocity in the $\theta$ direction $V_{\theta}^H$ is positive for $\theta \in (0,\pi/2)$ and negative for $\theta \in (\pi/2,\pi)$.

\subsection{The pair probability equation in spherical coordinates} \label{The pair probability equation in spherical coordinates}

In this subsection, we write the non-dimensional form of the steady-state pair probability equation in spherical coordinates. We scale the pair probability by $n_1n_2$ so that $P\in[0,1]$. We take $a^* = (a_1+a_2)/2$ and $\varGamma_{\eta} a^*$ as the characteristic length and velocity scale respectively. Thus, the non-dimensional radial separation between the centres of the two spheres, $r$ lies in the range of $2$ (referred to as the collision sphere) to $\infty$ (where one sphere does not influence the other). From here onward, we denote $r$ as the dimensionless relative distance between the centres. The size ratio $\kappa$, which can vary in the range $(0,1]$, captures the geometry of two-sphere system. The pair probability equation (\ref{Final_pair_probability_equation}) in spherical coordinate system becomes:
\begin{align}
    &\frac{1}{r^2}\frac{\partial}{\partial r}\Big[r^2\left(\mathcal{P}e V_r^g + V_r^H\right)P\Big]+\frac{1}{r\sin\theta}\frac{\partial}{\partial \theta}\Big[\left(\mathcal{P}e V_{\theta}^g + V_{\theta}^H\right) P\sin\theta\Big] -\nonumber \\
   &\frac{1}{r^2}\frac{\partial}{\partial r}\left[r^2 D^H_{rr} \frac{\partial P}{\partial r} + r D^H_{r\theta} \frac{\partial P}{\partial\theta}\right] -\frac{1}{r\sin\theta}\frac{\partial}{\partial \theta} \left[ \sin\theta \left(D^H_{r\theta} \frac{\partial P}{\partial r} + D^H_{\theta\theta} \frac{1}{r} \frac{\partial P}{\partial\theta} \right) \right]=0, 
\label{Non-dimensional_pair_probability_equation_in_spherical_coordinates}
\end{align}
where the dimensionless parameter $\mathcal{P}e$ measures the relative strength of advection induced by gravity to diffusion owing to fluctuations in the background flow. We label this parameter $\mathcal{P}e$ in the spirit of the P\'eclet number that appears in standard advection-diffusion equation. The expression for $\mathcal{P}e$ is as follows:
\begin{equation}
    \mathcal{P}e = \frac{Q Sv}{f\left(Re_{\lambda}\right)} = \frac{8\left(1-\kappa\right)}{81f\left(Re_{\lambda}\right)} \frac{Ar}{Fr}, 
\label{Definition_of_Pe}
\end{equation}
where $f(Re_{\lambda})$ is a function of the Taylor microscale Reynolds number (see Appendix \ref{appB} for the expression of $f\left(Re_{\lambda}\right)$), $Ar=\rho_p^2a_1^3g/\mu_f^2$ is the Archimedes number, and $Fr=St/Sv=\varGamma_{\eta}^2\eta/g$ is the Froude number. The Archimedes number depends on the particle and fluid properties and is independent of flow parameters, whereas the Froude number measures the competition of gravity with turbulence.

\begin{table}
\centering
\renewcommand{\arraystretch}{1.2}
\begin{tabular}{ccccccccc}
$\epsilon$
& $a_1$
& $St$
& $Sv$
& \multicolumn{5}{c}{$\mathcal{P}e$} \\
(m$^2$ s$^{-3}$)
& (\textmu m)
&
&
&
\multicolumn{5}{c}{$(Q)$} \\
\cline{5-9}
&
&
&
& $\kappa=0.99$
& $\kappa=0.9$
& $\kappa=0.7$
& $\kappa=0.5$
& $\kappa=0.3$ \\
\hline


$10^{-4}$
& $15$
& $0.016$
& $8.891$
& $2.776\times10^2$
& $2.776\times10^3$
& $8.328\times10^3$
& $1.388\times10^4$
& $1.943\times10^4$ \\

&
&
&
&
$(21.082)$
& $(2.108\times10^2)$
& $(6.325\times10^2)$
& $(1.054\times10^3)$
& $(1.476\times10^3)$ \\

\addlinespace

&
$20$
& $0.028$
& $15.811$
& $6.580\times10^2$
& $6.580\times10^3$
& $1.974\times10^4$
& $3.290\times10^4$
& $4.606\times10^4$ \\

&
&
&
&
$(28.109)$
& $(2.811\times10^2)$
& $(8.433\times10^2)$
& $(1.405\times10^3)$
& $(1.968\times10^3)$ \\

\addlinespace

&
$25$
& $0.044$
& $24.696$
& $1.285\times10^3$
& $1.285\times10^4$
& $3.855\times10^4$
& $6.426\times10^4$
& $8.996\times10^4$ \\

&
&
&
&
$(35.136)$
& $(3.514\times10^2)$
& $(1.054\times10^3)$
& $(1.757\times10^3)$
& $(2.460\times10^3)$ \\

\addlinespace

&
$30$
& $0.063$
& $35.566$
& $2.221\times10^3$
& $2.221\times10^4$
& $6.662\times10^4$
& $1.110\times10^5$
& $1.554\times10^5$ \\

&
&
&
&
$(42.164)$
& $(4.216\times10^2)$
& $(1.265\times10^3)$
& $(2.108\times10^3)$
& $(2.951\times10^3)$ \\

\hline


$5\times10^{-4}$
& $15$
& $0.035$
& $5.946$
& $83.018$
& $8.302\times10^2$
& $2.491\times10^3$
& $4.151\times10^3$
& $5.811\times10^3$ \\

&
&
&
&
$(9.428)$
& $(94.281)$
& $(2.828\times10^2)$
& $(4.714\times10^2)$
& $(6.600\times10^2)$ \\

\addlinespace

&
$20$
& $0.063$
& $10.566$
& $1.968\times10^2$
& $1.968\times10^3$
& $5.903\times10^3$
& $9.839\times10^3$
& $1.377\times10^4$ \\

&
&
&
&
$(12.571)$
& $(1.257\times10^2)$
& $(3.771\times10^2)$
& $(6.285\times10^2)$
& $(8.800\times10^2)$ \\

\addlinespace

&
$25$
& $0.098$
& $16.521$
& $3.843\times10^2$
& $3.843\times10^3$
& $1.153\times10^4$
& $1.922\times10^4$
& $2.690\times10^4$ \\

&
&
&
&
$(15.713)$
& $(1.571\times10^2)$
& $(4.714\times10^2)$
& $(7.857\times10^2)$
& $(1.100\times10^3)$ \\

\addlinespace

&
$30$
& $0.141$
& $23.784$
& $6.641\times10^2$
& $6.641\times10^3$
& $1.992\times10^4$
& $3.321\times10^4$
& $4.649\times10^4$ \\

&
&
&
&
$(18.856)$
& $(1.886\times10^2)$
& $(5.657\times10^2)$
& $(9.428\times10^2)$
& $(1.320\times10^3)$ \\

\hline


$3\times10^{-3}$
& $15$
& $0.087$
& $3.799$
& $21.655$
& $2.166\times10^2$
& $6.497\times10^2$
& $1.083\times10^3$
& $1.516\times10^3$ \\

&
&
&
&
$(3.849)$
& $(38.490)$
& $(1.155\times10^2)$
& $(1.925\times10^2)$
& $(2.694\times10^2)$ \\

\addlinespace

&
$20$
& $0.154$
& $6.754$
& $51.330$
& $5.133\times10^2$
& $1.540\times10^3$
& $2.567\times10^3$
& $3.593\times10^3$ \\

&
&
&
&
$(5.132)$
& $(51.320)$
& $(1.540\times10^2)$
& $(2.566\times10^2)$
& $(3.592\times10^2)$ \\

\addlinespace

&
$25$
& $0.241$
& $10.551$
& $1.003\times10^2$
& $1.003\times10^3$
& $3.008\times10^3$
& $5.013\times10^3$
& $7.018\times10^3$ \\

&
&
&
&
$(6.415)$
& $(64.150)$
& $(1.925\times10^2)$
& $(3.208\times10^2)$
& $(4.491\times10^2)$ \\

\addlinespace

&
$30$
& $0.346$
& $15.199$
& $1.732\times10^2$
& $1.732\times10^3$
& $5.197\times10^3$
& $8.662\times10^3$
& $1.213\times10^4$ \\

&
&
&
&
$(7.698)$
& $(76.980)$
& $(2.309\times10^2)$
& $(3.849\times10^2)$
& $(5.389\times10^2)$ \\

\hline
\end{tabular}

\caption{Representative values of the Stokes number $St$, settling parameter $Sv$, P\'eclet number $\mathcal{P}e$, and the corresponding parameter $Q$ for weakly turbulent warm-cloud conditions. For each combination of $\epsilon$, $a_1$, and $\kappa$, the values of $\mathcal{P}e$ are listed first, with the corresponding values of $Q$ given in parentheses directly beneath them.}
\label{tab:Pe_values}
\end{table}

Before proceeding further, we estimate the typical values of $\mathcal{P}e$ relevant to water droplets in warm clouds, where the turbulent dissipation rate typically lies in the range $\epsilon \sim 10^{-4}-10^{-1}$ m$^2$s$^{-3}$. Let us consider a pair of water droplets in a warm cumulus cloud with $a_1 = 10$ \textmu m, $\epsilon = 10^{-2}$ m$^2$s$^{-3}$, $\mu_f \approx 10^{-5}$ Pa s, $\nu_f \approx 10^{-5}$ m$^2$s$^{-1}$, and $\rho_p \approx 10^3$ kg m$^{-3}$. The corresponding $Ar$ and $Fr$ in this system are approximately $100$ and $0.056$, respectively. Using the model energy spectrum described in Appendix \ref{appB}, we obtain $f(Re_{\lambda})=0.675$ at $Re_{\lambda}=10^4$, which yields $\mathcal{P}e = 260.08(1-\kappa)$. Thus, for example, $\mathcal{P}e \approx 2.60$ for $\kappa=0.99$ and $\mathcal{P}e \approx 26.0$ for $\kappa=0.9$.

The parameter $Q$ can be expressed as $Q=(1-\kappa^2)Sv(\eta/a^*)$, indicating that, $Q$ is proportional to the settling parameter $Sv$. More generally, however, $Q$ also depends on the size ratio through $(1-\kappa^2)$ and on the ratio $\eta/a^*$. Consequently, $Q$ and $Sv$ can be varied independently by changing either the particle size ratio $\kappa$ or the ratio $\eta/a^*$ while keeping the remaining parameters fixed. This distinction is later exploited in the parameter study to isolate the separate influences of $Q$ and $Sv$ on the collision dynamics.

Table \ref{tab:Pe_values} summarizes representative values of $St$, $Sv$, $Q$, and $\mathcal{P}e$ over a range of turbulent dissipation rates, droplet radii, and size ratios relevant to weakly turbulent warm-cloud conditions. We consider $\epsilon$ in the range $10^{-4} - 3\times10^{-3}$ m$^2$s$^{-3}$ and droplet radii $a_1=15$--$30$ \textmu m, for which the rapid-settling limit ($St \ll 1$ and $Sv \gg 1$) can be simultaneously approached. Since $St \ll 1$ and $Sv \gg 1$ are asymptotic conditions, we use $St < 0.2$ and $Sv \geq 5$ as practical criteria. An important observation is that these two conditions are not independent. Since $Fr=St/Sv\propto\epsilon^{3/4}$, decreasing the turbulent dissipation rate simultaneously decreases $St$ and increases $Sv$. Consequently, the rapid-settling approximation becomes increasingly appropriate under weak-turbulence conditions. The tabulated values further show that, for a given $\epsilon$, $Sv$ increases with droplet radius, whereas increasing $\epsilon$ at a fixed droplet radius decreases $Sv$ and increases $St$. The P\'eclet number $\mathcal{P}e$ increases with increasing droplet size and size disparity and decreases with increasing turbulent dissipation rate, consistent with the scaling $\mathcal{P}e\propto a_1^3(1-\kappa)\epsilon^{-3/4}$. The corresponding values of $Q$ exhibit qualitatively similar trends. However, decreasing $\epsilon$, while improving the validity of the rapid-settling approximation, also increases $Q$ and $\mathcal{P}e$, thereby increasing the relative importance of differential settling compared with turbulent shear. Thus, at progressively weaker turbulence, the range of droplet sizes and size ratios for which turbulent shear makes an appreciable contribution to the collision dynamics shifts toward smaller and more nearly equal-sized droplets.

We assume that the particles stick on contact and form a permanent doublet (i.e., no rebound or subsequent break-up occurs). Therefore, the boundary condition at the contact surface is: $P = 0$ at $r = 2$. The absence of far-field correlations yields the other boundary condition: $P \rightarrow 1$ as $r \rightarrow \infty$. For the calculation purpose, we consider a large but finite value of the outer radius $r_{\infty}$ (where hydrodynamic interaction is insignificant, and beyond this point, the pair probability value has negligible variation). The axisymmetric geometry of two-sphere system allows us to analyse the problem in $\overline{\rho}-\overline{x}_3$ plane, where $\overline{x}_i=x_i/a^*$ ($i=1,2,3$) are dimensionless coordinates and $\overline{\rho}=(\overline{x}_1^2 + \overline{x}_2^2)^{1/2}$.

Instead of solving for pair probability equation, a popular approach for collision rate calculations in a laminar background flow is a trajectory analysis where the equations that need to be solved are entirely deterministic. There are studies that have incorporated the stochasticity due to thermal motion and solved an advection-diffusion equation either perturbatively (see \citealt{feke1983effect}) or numerically (see \citealt{zinchenko1994gravity,zinchenko1995collision}). Solving the pair probability equations for $\mathcal{P}e \gg 1$ is challenging because of boundary layer formations in the vicinity of the collision sphere. In the studies cited above, diffusivity and drift velocity are functions of radial separation only, whereas, in the present study, diffusivity and drift velocity are functions of both radial and angular coordinates and this increases the complexity of the problem.

\subsection{Expression for the collision rate}\label{Expression for the collision rate}

The radial pair probability flux integrated over the contact sphere defines the collision rate $K_{12}$. The flux due to radial relative velocity does not contribute to the collision rate since the pair probability vanishes at contact, and so the entire contribution comes from the radial component of the diffusive flux. Therefore, the collision rate normalized by $n_1n_2\varGamma_{\eta}\left(a_1+a_2\right)^3$ can be expressed as:
\begin{equation}
    \frac{K_{12}}{n_1n_2\varGamma_{\eta} \left(a_1+a_2\right)^3} = \frac{1}{8}\int_0^{2\pi}\int_0^{\pi}\Big[D^H_{rr} \frac{\partial P}{\partial r} + \frac{1}{r} D^H_{r\theta} \frac{\partial P}{\partial\theta}\Big]r^2\sin\theta d\theta d\phi|_{r=2}.
\label{Non-dimensional_collision_rate_basic_equation}
\end{equation}
After substituting the expressions of $D^H_{rr}$ and $D^H_{r\theta}$ from (\ref{Diffusivity_D_rr}) and (\ref{Diffusivity_D_rtheta}), the above equation reduces to:
\begin{equation}
    \frac{K_{12}}{n_1n_2\varGamma_{\eta} \left(a_1+a_2\right)^3} = 2\pi \frac{f(Re_{\lambda})}{Sv}\int_0^{\pi} F(\theta) \sin\theta d\theta,
\label{Non-dimensional_collision_rate}
\end{equation}
where 
\begin{align}
    F(\theta) &= (A-1)^2 \sin ^2\theta \left\{3 \cos 2 \theta +5 \right\} \left. \frac{\partial P}{\partial r}\right|_{r=2} \nonumber \\ &  + \frac{1}{4} (A-1) \sin 2 \theta \left\{ 3 (B-1) \cos 2 \theta +B+3 \right\}\left. \frac{\partial P}{\partial \theta}\right|_{r=2}.
\label{expression_of_F_theta}
\end{align}
We will show in the next section that the second term in (\ref{expression_of_F_theta}) is zero, and thus it does not contribute to the collision rate calculation. Therefore, the final expression for the collision rate becomes:
\begin{equation}
    \frac{K_{12}}{n_1n_2\varGamma_{\eta} \left(a_1+a_2\right)^3} = 2\pi \frac{f(Re_{\lambda})}{Sv}\int_0^{\pi} (A-1)^2 \sin ^3\theta \left\{3 \cos 2 \theta +5 \right\} \left. \frac{\partial P}{\partial r}\right|_{r=2}d\theta.
\label{Non-dimensional_collision_rate_final}
\end{equation}

\section{Semi-analytical solution for the pair probability} \label{Semi-analytical solution}

We exploit the axisymmetry of the present problem and express the solution for the pair-probability $P$ in terms of the following infinite series:
\begin{equation}
    P(r,\mu) = \sum_{n=0}^{\infty} a_n(r) \mathscr{P}_n(\mu), \label{Series_expansion_of_the_pair_probability}
\end{equation}
where $a_n(r)$ are the coefficients that depend only on the radial separation $r$, $\mathscr{P}_n(\mu)$ is the Legendre polynomial of order $n$, and $\mu = \cos\theta$. Substituting the above expansion for $P$ into the governing equation (\ref{Non-dimensional_pair_probability_equation_in_spherical_coordinates}) and then on integrating both side with $\mathscr{P}_m(\mu)$ over $\mu$ we obtain the following set of coupled ordinary differential equations (ODEs):
\begin{equation}
    \sum_{n=0}^{\infty} \mathcal{G}_n^m\frac{d^2a_n}{dr^2} + \left\{\frac{\mathcal{H}_n^m}{r} + \frac{\mathcal{P}e \hspace{1mm}\mathcal{I}_n^m}{r^2}\right\}\frac{da_n}{dr} + \left\{\frac{\mathcal{J}_n^m}{r^2} + \frac{\mathcal{P}e \hspace{1mm}\mathcal{N}_n^m}{r^3}\right\}a_n = 0. \label{Final_ode_for_the_pair_probability}
\end{equation}
The explicit expressions of the coefficients in (\ref{Final_ode_for_the_pair_probability}) are given in Appendix \ref{appC}. To obtain the pair probability, we need to find $a_n(r)$ by solving the system of coupled ODEs (\ref{Final_ode_for_the_pair_probability}) subjected to the following boundary conditions: 
\begin{equation}
    \left. \begin{array}{ll}
    a_n(r) = 0 \hspace{3mm} \textrm{at} \hspace{3mm} r = 2 \\[8pt]
    a_0(r) \rightarrow 1 \hspace{3mm}\textrm{as} \hspace{3mm} r \rightarrow \infty \\[8pt]
    a_n(r) \rightarrow 0 \hspace{3mm} \textrm{for} \hspace{3mm} n \neq 0 \hspace{3mm} \textrm{as} \hspace{3mm} r \rightarrow \infty.
    \end{array} \right\} \label{Boundary_conditions_in_terms_of_a_n}
\end{equation}
In the numerical simulations, we truncate the summation in (\ref{Final_ode_for_the_pair_probability}) at a finite number $N$ which introduce $N$ distinct variables ($a_0(r), a_1(r), \cdot\cdot\cdot, a_N(r)$) in the resulting expression. we can get $N$ such expressions by taking $m=0, 1, 2, \cdot\cdot\cdot, N$. In this way, a system of $N$ ODEs with each ODE involving $N$ variables is generated. We discretize each ODE in $r$ with fourth order accurate finite difference scheme. The grid points are non-uniformly spaced in $r$ with finer resolutions near the collision sphere in order to accurately capture the boundary layer at moderate to high $\mathcal{P}e$. We take logarithmically spaced $N_r$ points in the domain $2\leq r \leq r_{\infty}$. We can write the discretized system as $\boldsymbol{H}\cdot\boldsymbol{C}=\boldsymbol{R}$, where $\boldsymbol{C}$ is a $(N_r-2)N\times 1$ column vector containing the variables $a_n(r_i)$ on the different grid points with $(i = 2, 3, 4, \cdot\cdot\cdot, (N-1))$, $\boldsymbol{H}$  is a $N\times N$ block-matrix with the size of each block being $(N_r-2) \times (N_r-2)$. Each block matrix is tridiagonal for second order accurate finite difference scheme and is pentadiagonal for fourth order accurate finite difference scheme. The non-zero elements in the column matrix $\boldsymbol{R}$ arise from the non-homogeneous boundary conditions for $a_0(r)$ at far field. Finally, we solve the the linear system of equations using a direct LU-decomposition method taking the advantage of the sparse structure of $\boldsymbol{H}$. Previously, \cite{bergenholtz2002non,michelin2011optimal,shoele2018effects} have also adopted similar approach to solve for the advection-diffusion equation in spherical coordinates. This solution technique is commonly known as the Legendre polynomial spectral method (LPSM).

We can express the collision rate in terms of the coefficients $a_n$ by substituting (\ref{Series_expansion_of_the_pair_probability}) into (\ref{Non-dimensional_collision_rate_final}) and then evaluating the integral. Finally, we have:
\begin{equation}
    \frac{K_{12}}{n_1n_2\varGamma_{\eta} \left(a_1+a_2\right)^3} = \frac{f(Re_{\lambda})}{Sv}\frac{128\pi}{15}\left(A-1\right)^2 \left.\left[\frac{da_0}{dr}-\frac{1}{14}\left\{\frac{da_2}{dr}+ \frac{da_4}{dr}\right\}\right]\right|_{r=2}.
\end{equation}
Though the pair probability evaluation depends on all Legendre polynomial modes, the collision rate has explicit dependence on the first derivatives of three modes $a_0$, $a_2$, and $a_4$ at $r=2$. As per (\ref{Series_expansion_of_the_pair_probability}), the term involving $\partial P/\partial \theta|_{r=2}$ in (\ref{expression_of_F_theta}) depends on $a_n(2)$. From the boundary conditions in (\ref{Boundary_conditions_in_terms_of_a_n}), the coefficients $a_n(2)$ are zero and thus the term itself is zero.

\section{Ideal collision rate} \label{Ideal collision rate}

In the absence of hydrodynamic interactions, $A(r)=0$, $B(r)=0$, $L(r)=1$, and $M(r)=1$, which considerably simplifies the computation of the pair probability distribution and the corresponding ideal collision rate. Following the procedure described in \S\ref{Semi-analytical solution}, we compute the coefficients $a_n(r)$ by solving (\ref{Final_ode_for_the_pair_probability}) subject to the boundary conditions specified in (\ref{Boundary_conditions_in_terms_of_a_n}). We conduct numerical sensitivity tests to assess the convergence of our results, specifically examining their dependence on the number of radial grid points and the size of the computational domain. These tests confirm that $N_r = 500$ and $r_{\infty} = 100$ are sufficient to obtain well-converged results for $\mathcal{P}e \leq 10^3$. The sensitivity analysis further confirms that these values for $N_r$ and $r_{\infty}$ are adequate to resolve the boundary layer that forms near the collision surface in the same range of $\mathcal{P}e$. For the representative cloud parameters considered here, $r_{\infty}=100 a^*$ corresponds to a physical distance of approximately $1$ mm when $a^*=10$ \textmu m, which is comparable to the Kolmogorov length scale. Thus, the computational domain extends to the limit of the local linear-flow approximation while remaining within its validity range. The distribution of $P(r,\theta)$ exhibits a substantial dependence on the number of Legendre modes retained in (\ref{Series_expansion_of_the_pair_probability}) and (\ref{Final_ode_for_the_pair_probability}). To quantify this dependency on the number of modes $N$, we perform additional sensitivity analyses and find that $N=500$ is sufficient to accurately resolve the angular variation of $P(r,\theta)$ at the highest $\mathcal{P}e$ values considered in this study.

\begin{figure}
\centering
\includegraphics[width=1.0\textwidth]{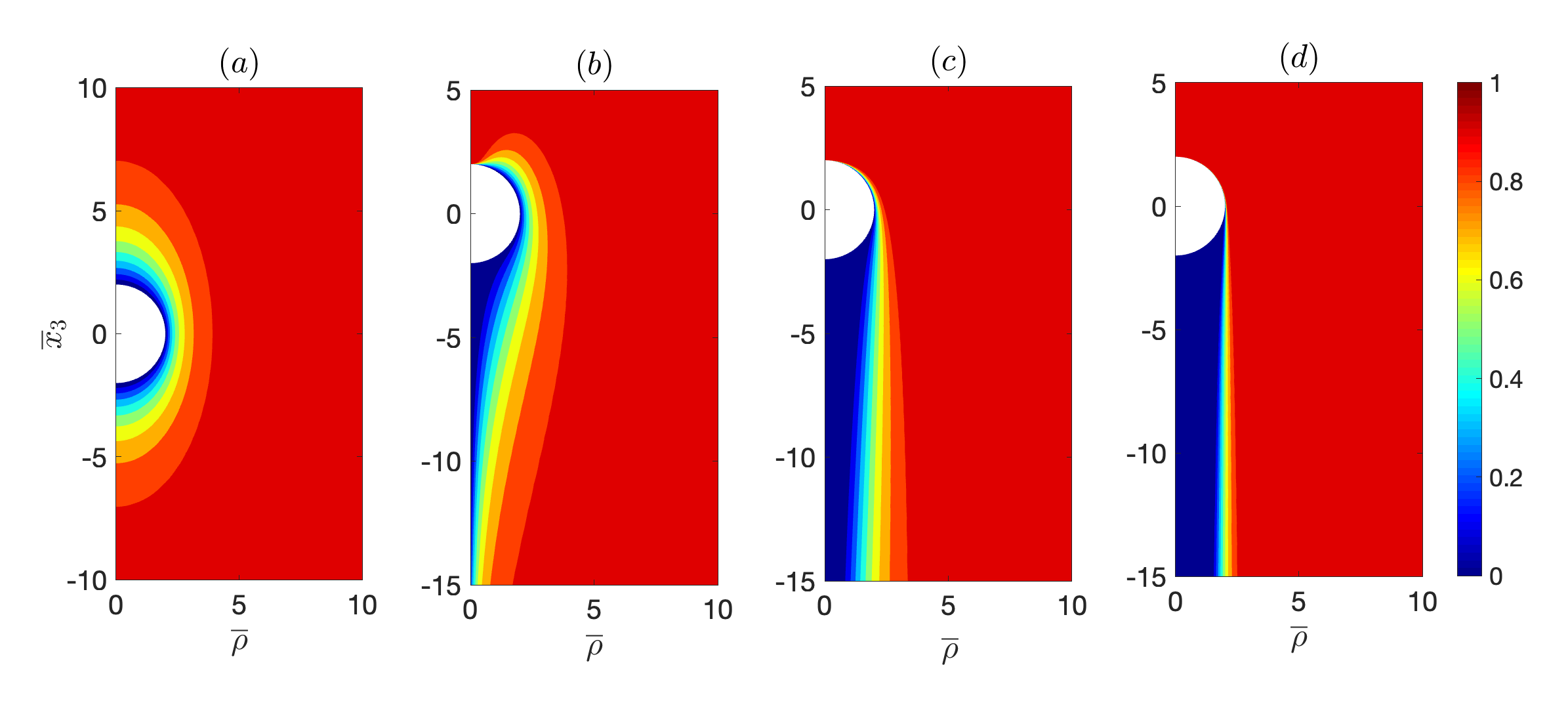}
\caption{Pair probability distribution for (a) $\mathcal{P}e = 0$, (b) $\mathcal{P}e = 10$, (c) $\mathcal{P}e = 10^2$, and (d) $\mathcal{P}e = 10^3$ in the absence of hydrodynamic interactions. For $\mathcal{P}e \rightarrow 0$, the pair-probability distribution is top-bottom symmetric. With increasing $\mathcal{P}e$, a wake region emerges in the bottom side of the collision sphere.}
\label{Ideal_collision_rate_Pair_probability}
\end{figure}

Figure (\ref{Ideal_collision_rate_Pair_probability}) shows the pair probability distribution around the collision surface for four distinct values of $\mathcal{P}e$ ($ = 0$, $10$, $10^2$, and $10^3$). At low $\mathcal{P}e$ values (typically when $\mathcal{P}e \ll 1$), the probability distribution is symmetric about the $\overline{x}_3=0$ line. In this regime, the diffusion term dominates over the advection term, and consequently, only the first few Legendre modes are sufficient to accurately compute $P(r,\theta)$. As $\mathcal{P}e$ increases, the top-bottom symmetry breaks, leading to an angular asymmetry in $P$. In particular, a probability-depleted wake region emerges on the bottom side of the collision sphere, while a region where $P \approx 1$ appears on the top side. With a further increase in $\mathcal{P}e$, the angular asymmetry becomes increasingly pronounced, characterized by a thin boundary layer on the upper side of the collision sphere and an extended wake region on the lower side. To accurately capture the wake region for $\mathcal{P}e \gg 1$, we use a large number of Legendre modes together with a refined radial grid resolution. At large $\mathcal{P}e$ values, the diffusion is insufficient to homogenize the strong angular gradients induced by advection, resulting in the formation of a wake in regions where the net relative radial velocity directs outward from the collision surface.

\begin{figure}
\centering
\includegraphics[width=1.0\textwidth]{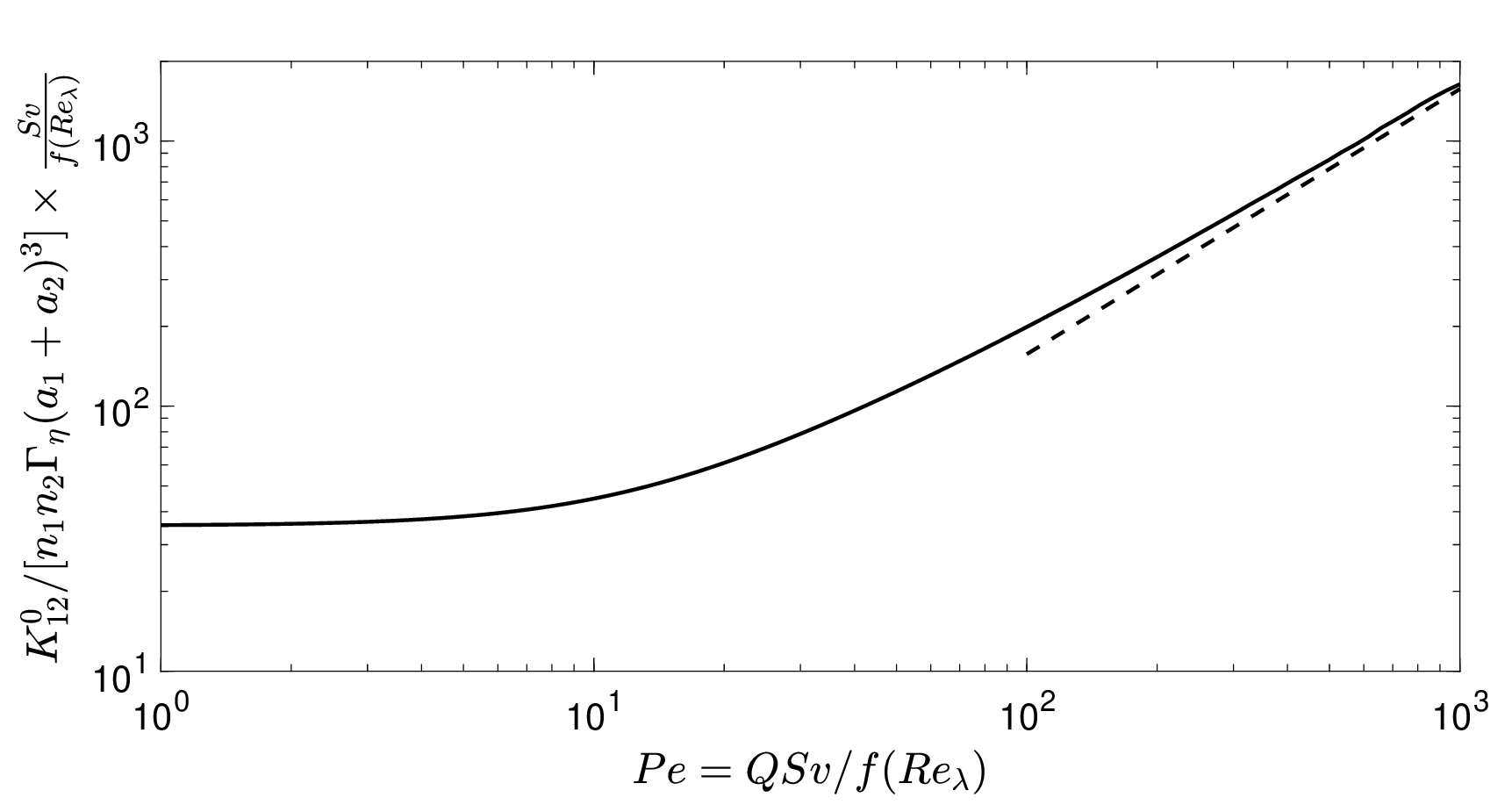}
\caption{The normalized ideal collision rate as a function of $\mathcal{P}e$. The collision rate increases monotonically with increasing $\mathcal{P}e$. The dashed line indicates the ideal collision rate due to pure differential sedimentation.}
\label{Ideal_collision_rate}
\end{figure}

\begin{figure}
\centering
\includegraphics[width=1.0\textwidth]{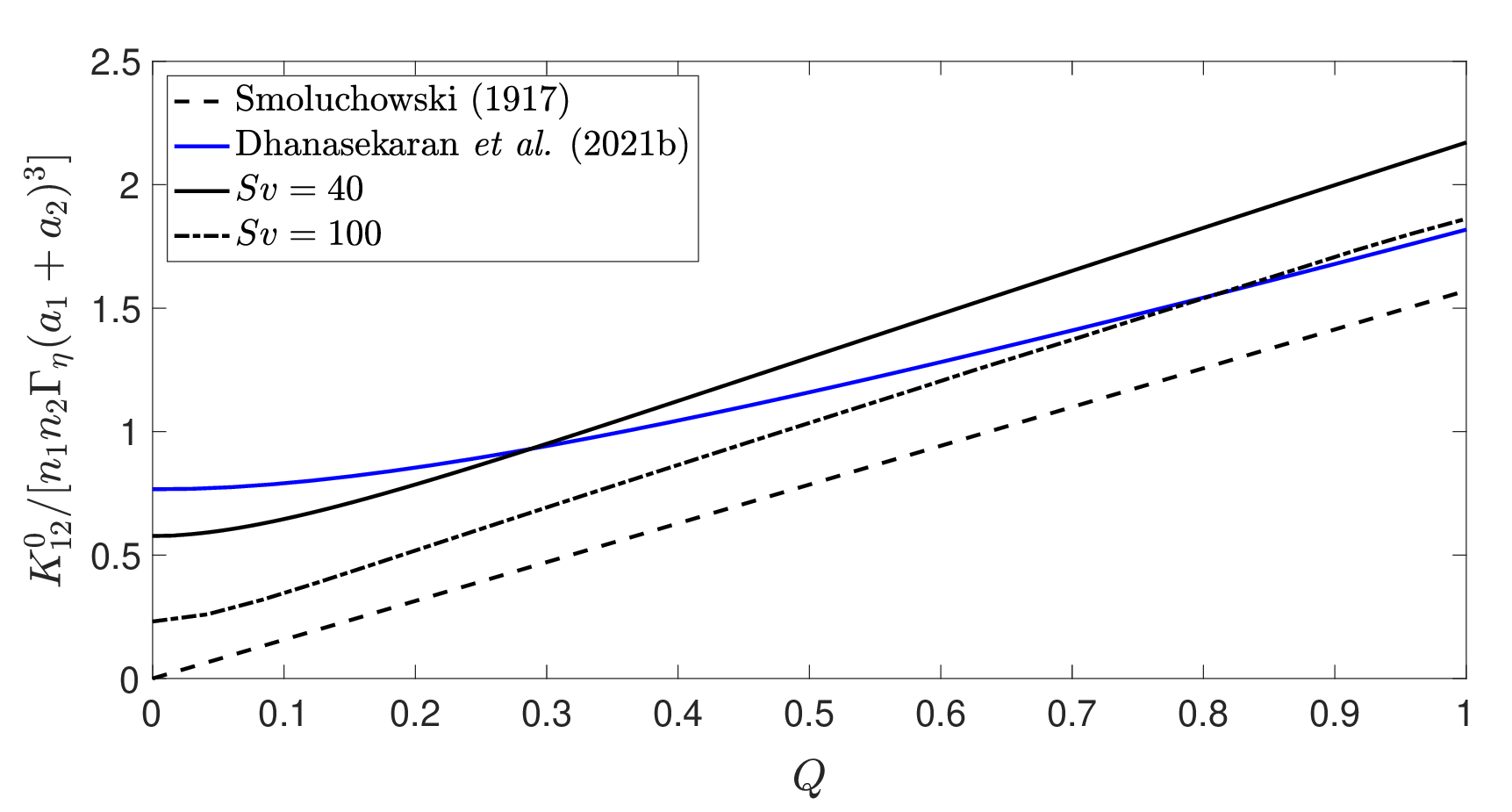}
\caption{The normalized ideal collision rate as a function of $Q$ for $Sv = 40, 100$ when $Re_{\lambda} = 2500$. We compare our results with the result for $Sv = 0$ obtained by \cite{dhanasekaran2021turbulent}.}
\label{Ideal_collision_rate_with_Q_for_Sv}
\end{figure}

\begin{figure}
\centering
\includegraphics[width=1.0\textwidth]{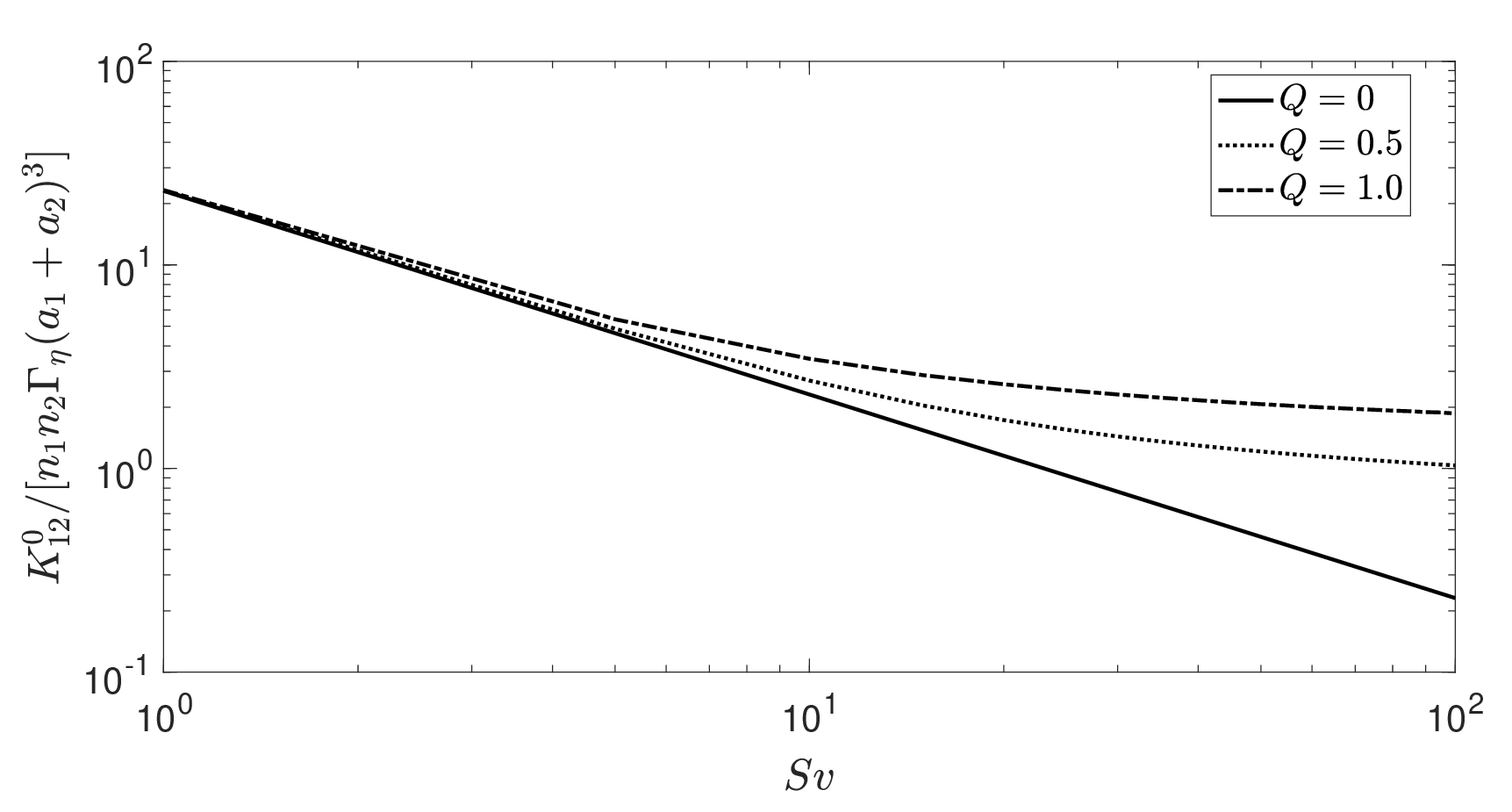}
\caption{The normalized ideal collision rate as a function of $Sv$ for different $Q$ when $Re_{\lambda}=2500$.}
\label{Ideal_collision_rate_with_Sv_for_different_Q}
\end{figure}

The ideal collision rate $K_{12}^0$, normalized by $n_1n_2 \varGamma_{\eta} \left(a_1+a_2\right)^3f(Re_{\lambda})/Sv$, is shown in figure \ref{Ideal_collision_rate}. For a given $Re_{\lambda}$ and $Sv$, the collision rate attains its minimum value at $\mathcal{P}e=0$. In the absence of gravity and hydrodynamic interactions, the advective terms vanish from the pair probability equation. Consequently, hydrodynamic diffusion experienced by rapidly settling sphere pairs in background turbulence solely drives the collision dynamics. As $\mathcal{P}e$ increases, the normalized collision rate increases monotonically and asymptotically approaches the Smoluchowski result for the collision rate of two differentially settling non-interacting spheres, shown by the dashed line in figure \ref{Ideal_collision_rate}.

Ignoring gravitational settling ($Q=0$), \cite{brunk1998turbulent} developed a simulation technique that computed the two-time Lagrangian statistics of the velocity gradient in an isotropic random flow and used trajectory analysis to calculate the coagulation rate for arbitrary total strain. For monodisperse particles, they showed how the ideal collision rate, normalized by particle radius and Kolmogorov shear rate, varies with total strain $\tau_S\varGamma_{\eta}$, while maintaining $\tau_S\varGamma_{\eta}/\tau_R\varGamma_{\eta}$ as $0.32$, corresponding to isotropic turbulence. In the small strain limit, using the pair diffusion formulation, \citet{brunk1997hydrodynamic} found that the normalized coagulation rate is $32 \pi (\varGamma_{\eta} \tau_S) /5$. At large total strain, \citet{brunk1998turbulent} reported a normalized coagulation rate of $9.896 \pm 0.0805$, whereas the model of \citet{saffman1956collision} predicts a value of $10.36$. For isotropic turbulence ($\tau_S\varGamma_{\eta}=2.3$ and $\tau_R\varGamma_{\eta}=7.2$), the normalized collision rate was found to be $8.62 \pm 0.02$ (see \citealt{brunk1997hydrodynamic}).  \cite{dhanasekaran2021turbulent} have recently examined the dependence of the ideal collision rate on $Re_{\lambda}$ for $Q=Sv=0$ and reported that the normalized collision rate decreases with increasing $Re_{\lambda}$, in contrast to the prediction of \citet{saffman1956collision}, who considered the rate independent of $Re_{\lambda}$. They also investigated the coupled effects of gravity and turbulence in the $Sv \rightarrow 0$ limit.

To compare our findings in the rapid-settling limit with those of \citet{dhanasekaran2021turbulent}, we first examine the scaling of the ideal collision rate. In this limit, the radial diffusivity scales as $D_{rr}\sim \varGamma_{\eta}^{2}a^{*2}\tau_v$, while the radial gradient of the pair probability at contact satisfies $\left.\partial P/\partial r\right|_{r=2}\sim 1/a^*$. Consequently, the collision rate scales as $n_1n_2(2a^*)^2 (\varGamma_{\eta}^2a^*{^2}\tau_v) (1/a^*) = n_1n_2\varGamma_{\eta }(a_1+a_2)^3/Sv$. This scaling analysis shows that, somewhat counterintuitively, the collision rate varies inversely with $Sv$ in the limit $Sv \gg 1$.

Keeping this scaling in mind, in figure \ref{Ideal_collision_rate_with_Q_for_Sv}, we show how the normalized collision rate $K_{12}^0/[n_1n_2 \varGamma_{\eta} \left(a_1+a_2\right)^3]$ varies with $Q$ for $Sv = 40$ and $100$ at $Re_{\lambda}=2500$. For a given $Sv$, the collision rate increases monotonically with $Q$. However, at small $Q$, the collision rate observed at higher $Sv$ (e.g., $Sv = 40$) is less than that in the $Sv \rightarrow 0$ limit. The collision rate decreases further at an even larger $Sv$ (for example, $Sv = 100$). Physically, rapid settling reduces the effective turbulent strain experienced by the particle pairs, resulting in a lower collision rate compared to the $Sv \rightarrow 0$ case. At higher $Q$, however, rapid settling enhances the gravitational contribution to the collision process, leading to collision rates higher than those obtained in the $Sv \rightarrow 0$ regime.

Figure \ref{Ideal_collision_rate_with_Sv_for_different_Q} shows the variation of the ideal collision rate with $Sv$ for $Q=0$, $0.5$, and $1.0$ at $Re_{\lambda}=2500$. Consistent with the scaling analysis presented above, the collision rate decreases with increasing $Sv$ for small but finite values of $Q$. In the limiting case of purely turbulent-shear-induced collisions (i.e., when $Q=0$), the collision rate is inversely proportional to $Sv$. As $Q$ increases, the gravitational contribution to the relative motion becomes progressively more significant. Consequently, at sufficiently large $Sv$, the collision rate increases with increasing $Q$, reflecting the increasingly dominant role of differential settling in the collision dynamics.

\section{Collision with hydrodynamic interactions} \label{Collision with hydrodynamic interactions}

\begin{figure}
\centering
\includegraphics[width=1.0\textwidth]{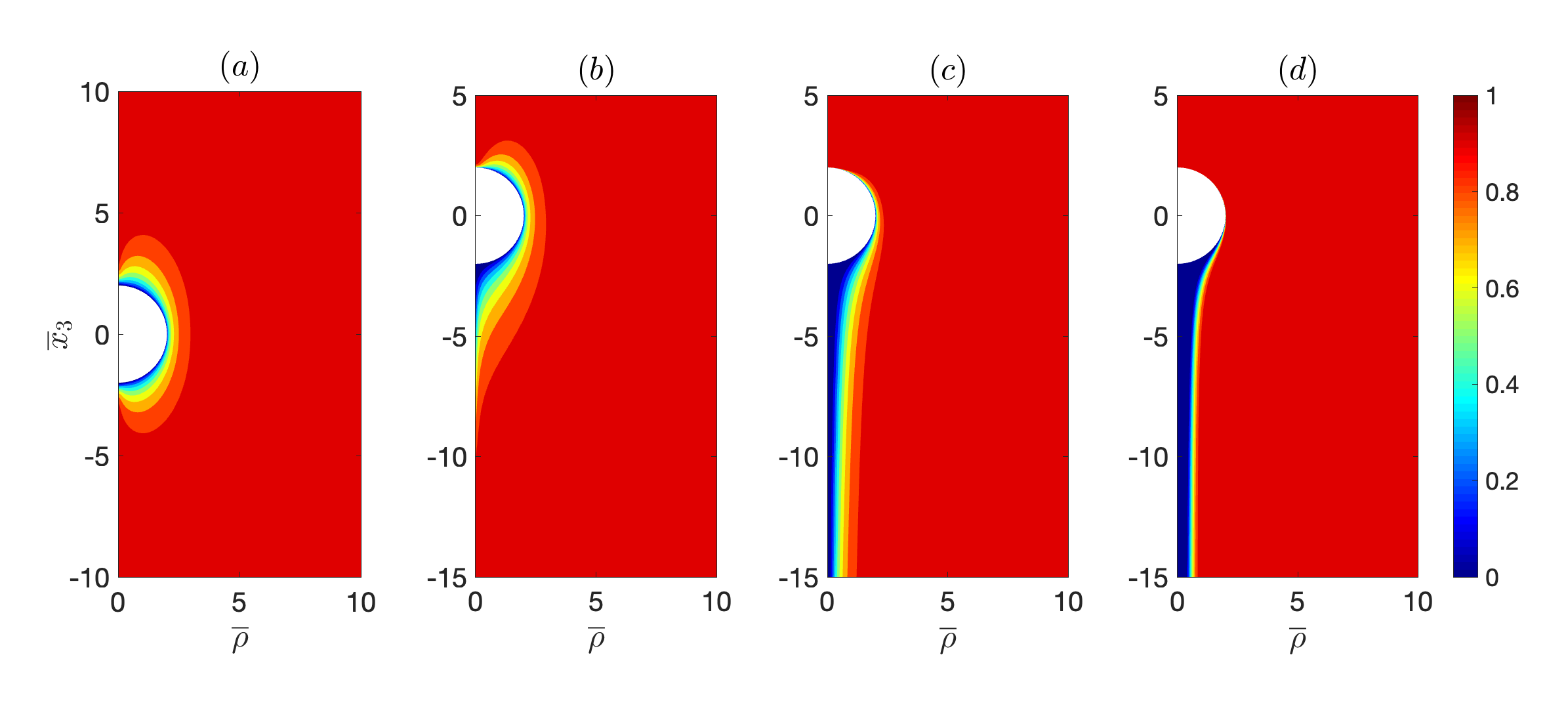}
\caption{Pair probability with hydrodynamic interactions in the $(r,\theta)$ plane for $(a) \mathcal{P}e = 0$, $(b) \mathcal{P}e = 10$, $(c) \mathcal{P}e = 10^2$, and $(d) \mathcal{P}e = 10^3$ when $\kappa=0.99$ and $Kn=10^{-1}$.}
\label{probability_contour_kappa_099_Kn_01}
\end{figure}

Hydrodynamic interactions modify both the pair diffusivity and relative particle velocity, thereby altering the pair probability distribution. The collision rate that accounts for hydrodynamic interactions, $K_{12}$, is generally lower than the corresponding ideal collision rate $K_{12}^0$. To quantify the reduction in collision rate caused by hydrodynamic interactions, we define the collision efficiency as $E_{12}= K_{12}/K_{12}^0$. Non-continuum lubrication interactions introduce an additional dependence of the collision dynamics on the Knudsen number $Kn$. Consequently, in addition to $\mathcal{P}e$, the collision efficiency also depends on $Kn$ and $\kappa$. In this study, we present key results for two representative values of the Knudsen number: $Kn=10^{-2}$, where, as two drops approach one another, a substantial range of near-continuum lubrication precedes the onset of non-continuum lubrication effects, and $Kn=10^{-1}$, for which non-continuum effects modify the lubrication interactions over their full range. Similarly, we consider two representative particle size ratios, $\kappa=0.6$, corresponding to a noticeably bidisperse pair, and $\kappa=0.99$, representing an almost equal-sized pair. For each combination of $Kn$ and $\kappa$, the turbulent Peclet number $\mathcal{P}e$ is varied over the range $0 \leq \mathcal{P}e \leq 10^3$.

As two spheres approach one another, the strong divergence of continuum lubrication resistance prevents surface-to-surface contact from occurring in a finite time under the action of a constant force. In contrast, the weaker divergence of the non-continuum lubrication resistance allows collisions to occur in finite time. Non-continuum effects become important when $s=r-2=O(Kn)$. For continuum lubrication interactions, the radial mobilities $(1-A)$, $L$ scale as $O(s)$, whereas for non-continuum lubrication interactions these mobilities scale as $O(1/\ln[\ln(Kn/s)])$ (see \citealt{sundar96non}). However, exactly at $s=0$, the radial relative velocities vanish even in the presence of non-continuum effects where $A=1$ and $L=0$. To avoid numerical difficulties in pair trajectory in the presence of this weak integrable singularity, we consider an offset collision sphere of radius $r=2+s_0$. We present converged results for $s_0=10^{-6}$.

\begin{figure}
\centering
\includegraphics[width=1.0\textwidth]{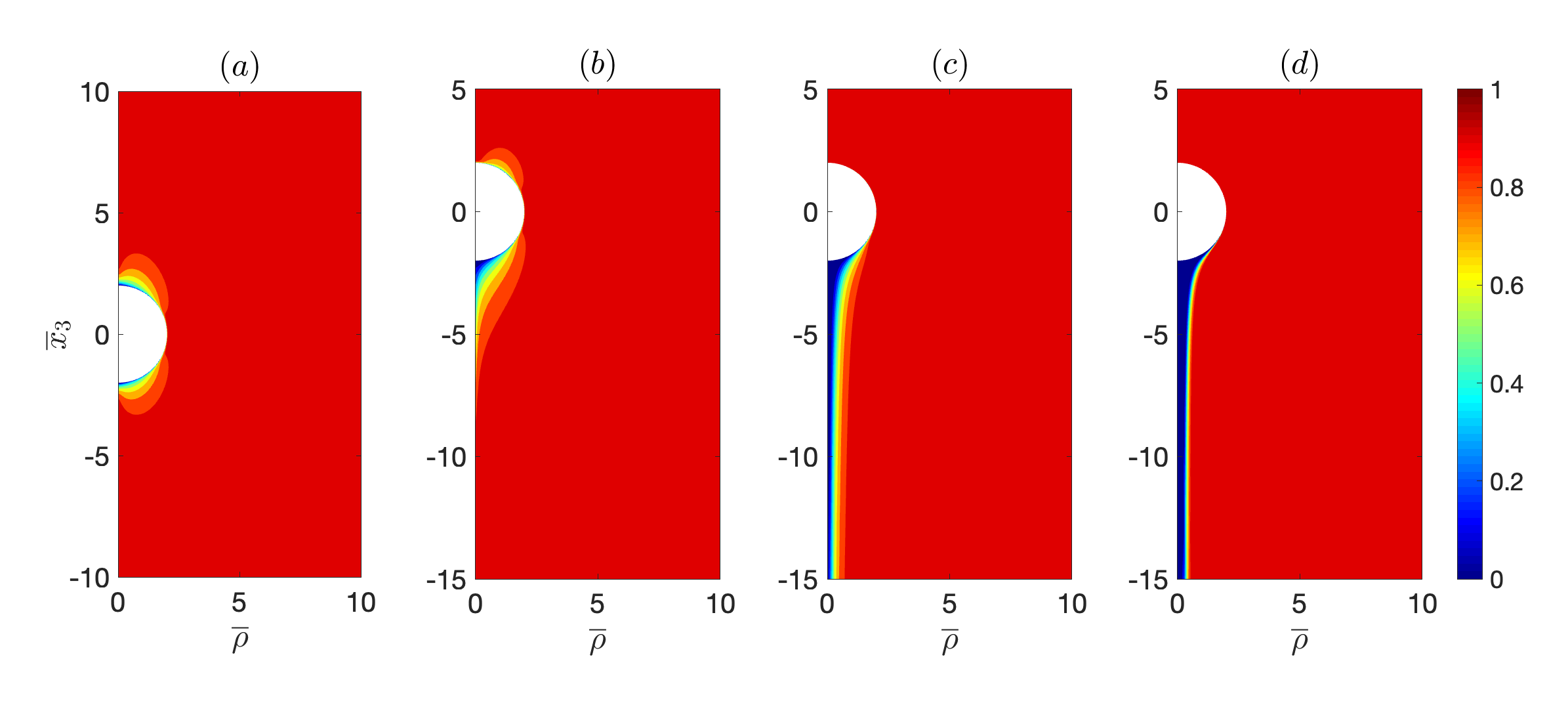}
\caption{Pair probability with hydrodynamic interactions in the $(r,\theta)$ plane for $(a) \mathcal{P}e = 0$, $(b) \mathcal{P}e = 10$, $(c) \mathcal{P}e = 10^2$, $(d) \mathcal{P}e = 10^3$ when $\kappa=0.99$ and $Kn=10^{-2}$.}
\label{probability_contour_kappa_099_Kn_001}
\end{figure}

\begin{figure}
\centering
\includegraphics[width=1.0\textwidth]{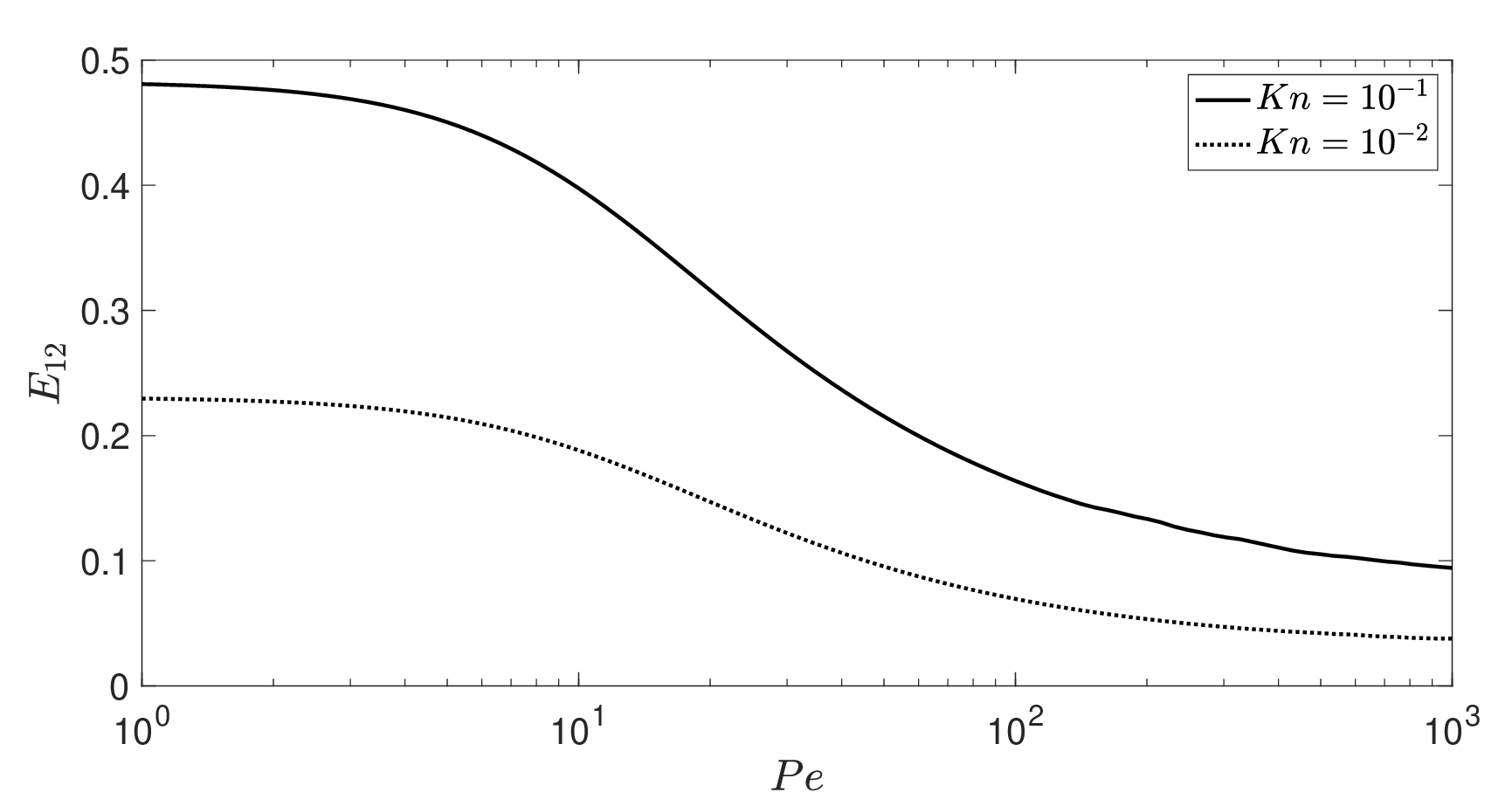}
\caption{The collision efficiency as a function of $\mathcal{P}e$ for $Kn = 10^{-1}, 10^{-2}$ when $\kappa = 0.99$.}
\label{Collision_efficiency_kappa_099}
\end{figure}

\begin{figure}
\centering
\includegraphics[width=1.0\textwidth]{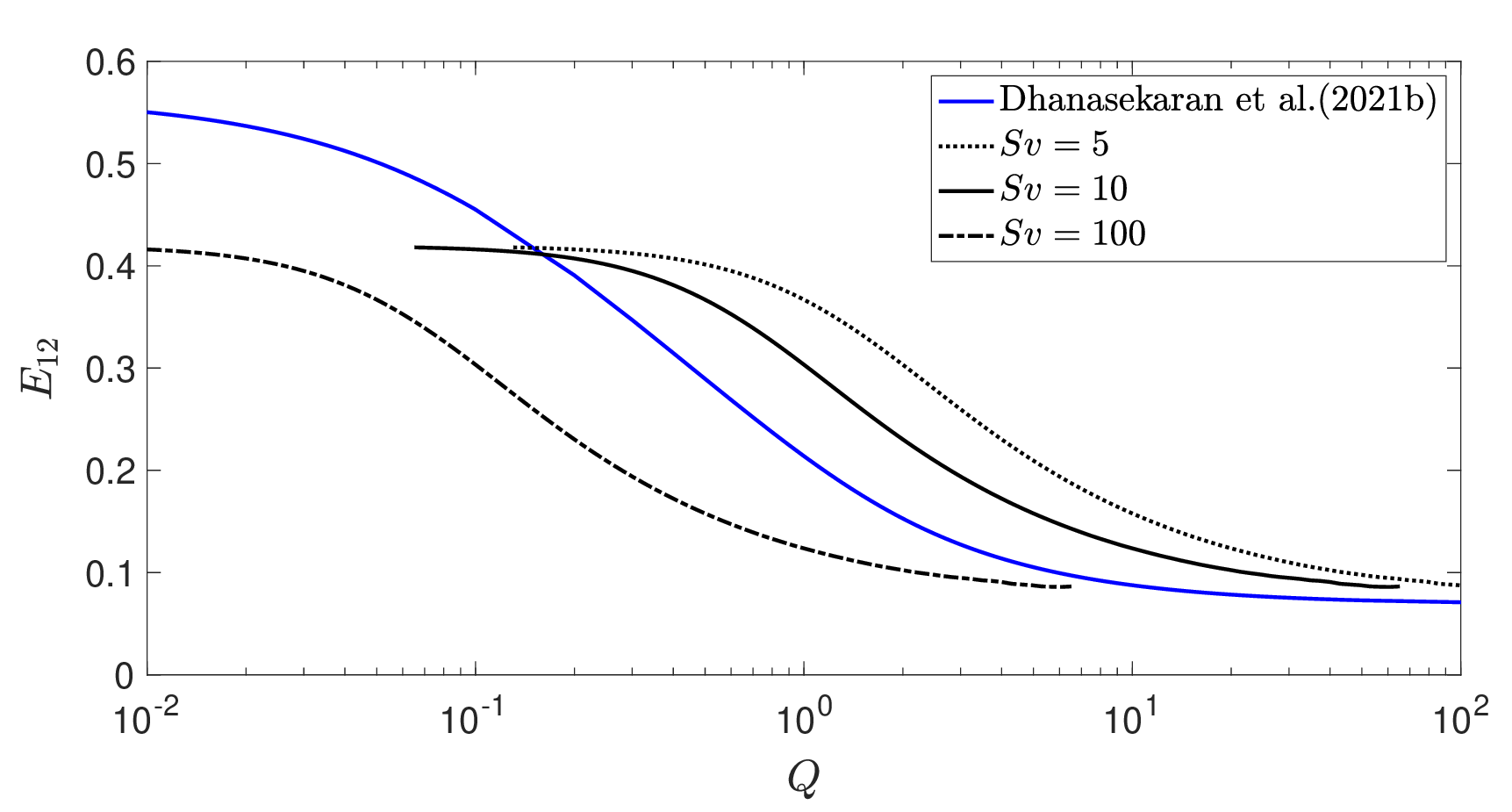}
\caption{The collision efficiency as a function of $Q$ for different values of $Sv$ when $Kn = 10^{-1}$, $\kappa = 0.6$ and $Re_{\lambda}=2500$. The continuous blue line is the results for $Sv = 0$ from \cite{dhanasekaran2021turbulent}.}
\label{Collision_efficiency_kappa_06}
\end{figure}

\begin{figure}
\centering
\includegraphics[width=1.0\textwidth]{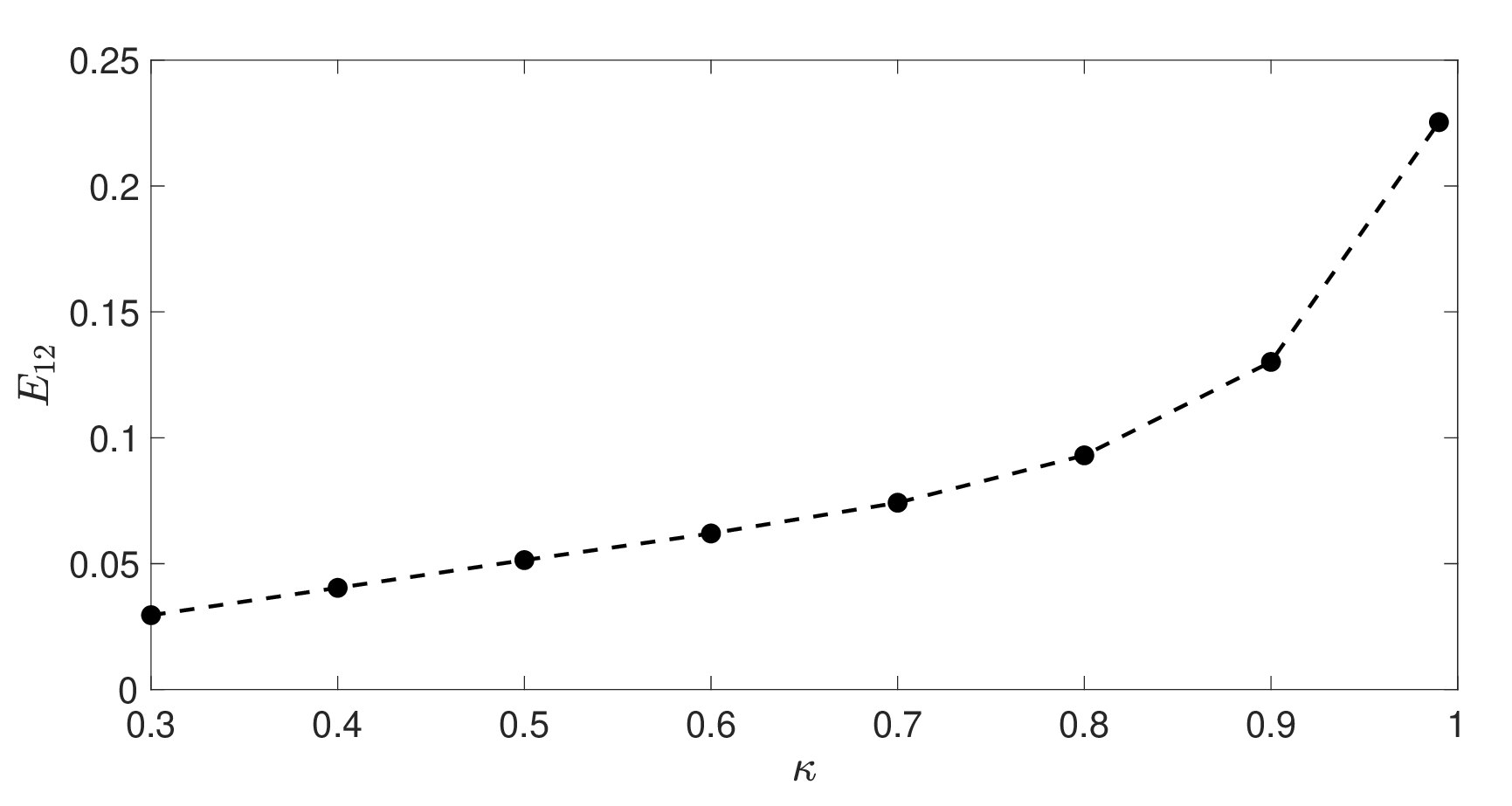}
\caption{The collision efficiency as a function of $\kappa$ with $a_1 = 10$ \textmu m for a cumulus clouds with $\epsilon = 10^{-2}$ m$^2$s$^{-3}$ and $Re_{\lambda}=2500$.}
\label{Collision_efficiency_for_a1_10}
\end{figure}

We follow the methodology outlined in \S\ref{Semi-analytical solution} to compute the steady-state pair probability distribution while accounting for hydrodynamic interactions. In this case, accurately capturing the wake structure requires a significantly larger number of Legendre modes, even for smaller values of $\mathcal{P}e$. Numerical sensitivity tests indicate that approximately $N \approx 50$ modes are sufficient for $\mathcal{P}e = 1$, whereas $N \approx 10^3$ is adequate for $\mathcal{P}e = 10^3$. However, akin to the ideal case, we find that $N_r = 500$ and $r_{\infty} = 100$ are sufficient to obtain converged results.

Figures \ref{probability_contour_kappa_099_Kn_01} and \ref{probability_contour_kappa_099_Kn_001} show the distribution of $P(r,\theta)$ for four distinct values of $\mathcal{P}e$, with $\kappa=0.99$ and $Kn=10^{-1}$ and $Kn=10^{-2}$, respectively. As expected, for both values of $Kn$, the probability distributions exhibit top-bottom symmetry when $\mathcal{P}e \ll 1$. However, the distributions corresponding to $Kn=10^{-1}$ and $Kn=10^{-2}$ differ substantially from one another. More importantly, the distributions near $\theta=0$ and $\theta=\pi$ show substantial deviations from the ideal case discussed in \S\ref{Ideal collision rate}. For $Kn=10^{-2}$, two distinct but symmetric regions appear on the upper and lower sides of the contact sphere. As $\mathcal{P}e$ increases, a boundary layer forms on the upper side of the sphere, while a wake region emerges on the lower side. With further increases in $\mathcal{P}e$, the boundary layer becomes progressively thinner, and the wake region becomes increasingly wider and longer. These wake regions are substantially smaller than those observed in the absence of hydrodynamic interactions. Furthermore, for a fixed $\mathcal{P}e$ and $\kappa$, the wake region corresponding to $Kn=10^{-2}$ is considerably thinner than that for $Kn=10^{-1}$ because fewer collisions occur with relatively weak non-continuum lubrication effects.

Figure \ref{Collision_efficiency_kappa_099} shows the variation of the collision efficiency with $\mathcal{P}e$ for $\kappa=0.99$ and $Kn=10^{-1}, 10^{-2}$. The collision efficiency decreases monotonically with increasing $\mathcal{P}e$, attaining its maximum value in the turbulence-dominated regime (i.e., as $\mathcal{P}e \rightarrow 0$) and its minimum value in the gravity-dominated regime (i.e., as $\mathcal{P}e \rightarrow \infty$). As $Kn$ decreases, the range of separation distances over which continuum lubrication resistance acts increases, resulting in stronger hydrodynamic resistance to the radial approach of the particle pair. Consequently, the collision efficiency for $Kn=10^{-2}$ is substantially smaller than that for $Kn=10^{-1}$ over the entire range of $\mathcal{P}e$ considered here. For two cloud droplets of fixed sizes, $\mathcal{P}e$ increases as the turbulent dissipation rate decreases (see Table \ref{tab:Pe_values}). The present results, therefore, imply that droplet collisions are more efficient in highly turbulent clouds, where turbulence-induced relative motion is less strongly affected by lubrication resistance than differential-settling-induced relative motion. As the dynamics become increasingly gravity-dominated (i.e., with increasing $\mathcal{P}e$), lubrication interactions suppress radial approach more effectively, resulting in lower collision efficiency.

In figure \ref{Collision_efficiency_kappa_06}, we present the variation of the collision efficiency with the relative strength of gravity and turbulence for $Sv = 5, 10,$ and $100$, with $Kn=10^{-1}$, $\kappa = 0.6$, and $Re_{\lambda}=2500$. Consistent with the results in the $Sv \rightarrow 0$ limit reported by \citet{dhanasekaran2021turbulent}, the collision efficiency decreases monotonically with increasing $Q$. Furthermore, for any fixed value of $Q$, the collision efficiency predicted by the present theory decreases monotonically with increasing $Sv$. This behavior follows directly from the relation $\mathcal{P}e = QSv/f(Re_{\lambda})$, together with the monotonic decrease of the collision efficiency with increasing $\mathcal{P}e$. However, a comparison with the $Sv \rightarrow 0$ results should be interpreted with caution. The $Sv \rightarrow 0$ limit corresponds to a fundamentally different asymptotic regime, in which particle pairs sample the velocity-gradient field along fluid trajectories over a finite total strain of order unity (specifically, $\tau_S\Gamma_{\eta} \approx 2.3$ for homogeneous isotropic turbulence). In contrast, the present theory assumes the rapid settling limit ($Sv \gg 1$), where the total strain experienced by a particle pair is asymptotically small ($\Gamma_{\eta}\tau_v \ll 1$). Consequently, the two asymptotic formulations are not continuously connected within the small total strain framework adopted here, and the apparent crossing between the $Sv \rightarrow 0$ and finite-$Sv$ results reflects the change in the underlying mechanism governing the turbulence-induced relative motion, rather than a genuine non-monotonic dependence of the collision efficiency on $Sv$.

To examine the role of polydispersity on the collision dynamics, we investigate the collision efficiency as a function of $\kappa$ for a representative cumulus cloud condition with $\epsilon = 10^{-2}$ m$^2$s$^{-3}$ and $Re_{\lambda}=2500$, while fixing $a_1 = 10$~\textmu m (see figure \ref{Collision_efficiency_for_a1_10}). The collision efficiency increases with increasing particle size ratio $\kappa$ because nearly equal-sized particles experience weaker differential settling, thereby reducing the gravity-driven component of the radial approach that is most strongly affected by lubrication resistance. Consequently, the relative contribution of turbulence-induced motion becomes more significant, leading to higher collision efficiencies. In contrast, highly bidisperse particle pairs experience stronger differential settling, so that lubrication interactions suppress the radial approach more effectively, resulting in a lower collision efficiency.

\section{Summary and conclusions} \label{Summary and conclusions}

We have studied the collision dynamics of hydrodynamically interacting inertialess spherical particle pairs settling rapidly through homogeneous isotropic turbulence. The analysis focused on the asymptotic regime $Sv \gg 1$ and $St \ll 1$, which is relevant to sub-Kolmogorov droplets in atmospheric clouds where gravitational settling dominates turbulent velocity fluctuations while particle inertia remains negligible. Our study includes hydrodynamic interactions between particles, accounting for the breakdown of continuum approximations at close separations. The non-continuum lubrication resistance exhibits a weaker divergence compared to its continuum counterpart, thereby facilitating surface-to-surface contact in finite time.

In the rapid-settling limit, a pair of particles traverses a Kolmogorov eddy in a time much shorter than the eddy turnover time. Consequently, the turbulent velocity gradients sampled along the settling trajectory decorrelate rapidly, resulting in a total strain accumulated during one correlation time that scales as $Sv^{-1} \ll 1$. In this scenario, the relative motion induced by turbulent shear can be effectively modeled as a diffusion process. Employing a framework analogous to the diffusion theory developed by \citet{kesten1979limit}, we have derived a Fokker–Planck equation that governs the probability density function of the particle pair. The resulting equation contains a hydrodynamic diffusivity tensor and a relative drift velocity arising from the nonsolenoidal nature of the hydrodynamic-interaction-induced relative velocity.

A key aspect of the analysis involved the evaluation of the velocity-gradient autocorrelation function along the settling trajectory. Exploiting the rapid settling approximation, the Lagrangian autocorrelation function was related directly to the turbulence energy spectrum. Using a model energy spectrum for homogeneous isotropic turbulence, explicit expressions for the diffusivity coefficients were obtained. The resulting diffusivity scales inversely with the settling parameter, reflecting the reduced influence of turbulent fluctuations as the settling velocity increases.

We solved the pair probability equation semi-analytically by exploiting the axisymmetric nature of the problem. In the absence of hydrodynamic interactions, the collision rate depends on the relative strength of gravity to turbulence ($\mathcal{P}e$), the Taylor microscale Reynolds number ($Re_{\lambda}$), and the settling parameter ($Sv$). We have shown that, for a given $Re_{\lambda}$, the ideal collision rate increases monotonically with increasing relative strength of gravity compared with turbulence. Physically, stronger differential sedimentation enhances the radial flux of approaching particle pairs, thereby increasing the collision rate. However, hydrodynamic interactions substantially impede close particle approach through lubrication resistance and introduce strong dependence on $Kn$ and $\kappa$. As a result, although the ideal collision rate increases, the collision efficiency decreases monotonically with increasing gravitational influence.

Although the current study is restricted to $St \ll 1$ and the absence of non-hydrodynamic forces, the framework may be extended to include finite particle inertia and non-hydrodynamic interparticle interactions such as electrostatic interactions. Such extensions are expected to be more relevant to accurately predicting the collision rates of polydisperse droplets in the larger portion of the $15-40$ \textmu m radius range associated with the condensation–coalescence bottleneck in warm cloud microphysics. More broadly, the framework developed here may serve as a basis for improved collision kernels in cloud microphysical parameterizations.

\vspace{5mm}

\textbf{Acknowledgements.} AR acknowledges the financial support received from the Anusandhan National Research Foundation (ANRF), Government of India, through the Advanced Research Grant (ANRF/ARG/2025/010666/ENS). PP would like to acknowledge financial support from the Prime Minister’s Research Fellows (PMRF) scheme, Ministry of Education, Government of India (Project Number: SB22230184AMPMRF008746). DLK acknowledges support from National Science Foundation grant 2535828. 

\vspace{5mm}
\textbf{Declaration of interest.} The authors report no conflict of interest.

\appendix
\section{The turbulence energy spectrum model and the expression for $\alpha_2$}\label{appB}
We adopt the model energy spectrum for homogeneous isotropic turbulence given in \citet{pope_2000} (see pp. 232--233), expressed as
\begin{equation}
    E(\xi) = C\epsilon^{2/3}\xi^{-5/3}f_L(\xi L)f_{\eta}(\xi\eta), \label{Turbulenve_energy_spectrum}
\end{equation}
where
\begin{align}
    f_L(\xi L) &= \left[\frac{\xi L}{\sqrt{(\xi L)^2 + c_L}}\right]^{5/3 +p_0}, \label{Expression_of_f_L}\\
    f_{\eta}(\xi\eta) &= \exp\left[-\beta\left \{\left[(\xi\eta)^4 + c_{\eta}^4 \right]^{1/4} - c_{\eta}\right\}\right]. \label{Expression_of_f_eta}
\end{align}
The values of the parameters involved in the above three expressions are: $C=1.5$, $\beta=5.2$, $p_0=2.0$, $c_L =6.78$ and $c_{\eta}=0.40$. Substituting \eqref{Turbulenve_energy_spectrum} into \eqref{General_expressions_of_alpha1_and_alpha2}, the coefficient $\alpha_2$ can be expressed as
\begin{equation}
    \alpha_2 = -\frac{\pi C\epsilon^{2/3} L^{-1/3}}{16g\tau_p}\int_{0}^{\infty} dz \exp\left[-\beta\left\{\left[(z\zeta)^4 + c_{\eta}^4 \right]^{1/4} - c_{\eta}\right\}\right] \left[\frac{z}{\sqrt{z^2 + c_L}}\right]^{5/3 +p_0} z^{-2/3}, \label{Full_expression_of_alpha_2}
\end{equation}
where $\zeta = \eta /L = (3 Re_{\lambda}^2 /20)^{-3/4}$. Using the relations $Sv = g\tau_p/u_{\eta}$, equation \eqref{Full_expression_of_alpha_2} reduces to
\begin{equation}
   \alpha_2 = -\frac{\pi C\zeta^{1/3}}{16 Sv \tau_{\eta}}\int_{0}^{\infty} dz \exp\left[-\beta\left\{ \left[(z\zeta)^4 + c_{\eta}^4\right]^{1/4} - c_{\eta}\right\} \right] \left[\frac{z}{\sqrt{z^2 + c_L}}\right]^{5/3 +p_0} z^{-2/3}. \label{Simplified_expression_of_alpha_2} 
\end{equation}
It is convenient to express $\alpha_2$ in the form $\alpha_2 = -f(Re_{\lambda})\varGamma_{\eta}/Sv$, where the $f(Re_{\lambda})$ is given by:
\begin{equation}
    f(Re_{\lambda}) = \frac{\pi C\zeta^{1/3}}{16} \int_{0}^{\infty}dz \exp\left[-\beta\left\{\left[(z\zeta)^4 + c_{\eta}^4\right]^{1/4} - c_{\eta}\right\} \right] \left[\frac{z}{\sqrt{z^2 + c_L}}\right]^{5/3 +p_0} z^{-2/3}. \label{expresiion_f(Re_lambda)}
\end{equation}

The asymptotic behaviour of $f(Re_{\lambda})$ for large $Re_{\lambda}$ can be inferred from the structure of the integral in \eqref{expresiion_f(Re_lambda)}. In the inertial subrange, where $1 \ll z \ll \zeta^{-1}$, the dissipative cutoff function satisfies $f_{\eta}\approx 1$, while the large-scale correction satisfies $z \gg \sqrt{c_L}$, so that
\[
\left[\frac{z}{\sqrt{z^2+c_L}}\right]^{5/3+p_0}\approx 1.
\]
Consequently, the integrand behaves asymptotically as $z^{-2/3}$, and the dominant contribution from the inertial subrange scales as
\[
\zeta^{1/3}\int^{O(\zeta^{-1})}z^{-2/3}\,dz
\sim
\zeta^{1/3}(\zeta^{-1})^{1/3}
=O(1),
\]
which explains why $f(Re_{\lambda})$ approaches a finite constant as $Re_{\lambda}\rightarrow\infty$. The leading finite-Reynolds-number correction arises from the energy-containing and dissipation ranges, whose contributions remain $O(1)$ in the integral and are therefore multiplied only by the prefactor $\zeta^{1/3}$. Since
\[
\zeta^{1/3} = \left(\frac{3Re_{\lambda}^{2}}{20}\right)^{-1/4} \propto Re_{\lambda}^{-1/2},
\]
the large-$Re_{\lambda}$ behaviour is expected to be
\begin{equation}
    f(Re_{\lambda}) \approx f_{\infty} - k\,Re_{\lambda}^{-1/2},
    \label{asymptotic_fRelam}
\end{equation}
where $f_{\infty}$ is the asymptotic value of $f(Re_{\lambda})$ as $Re_{\lambda} \rightarrow \infty$, and $k$ is a positive constant. By fitting the numerical evaluation of \eqref{expresiion_f(Re_lambda)} at sufficiently large values of $Re_{\lambda}$, we find that $f_{\infty} \approx 0.698$ and $k \approx 2.34$.

\begin{figure}
\centering
\includegraphics[width=1.0\textwidth]{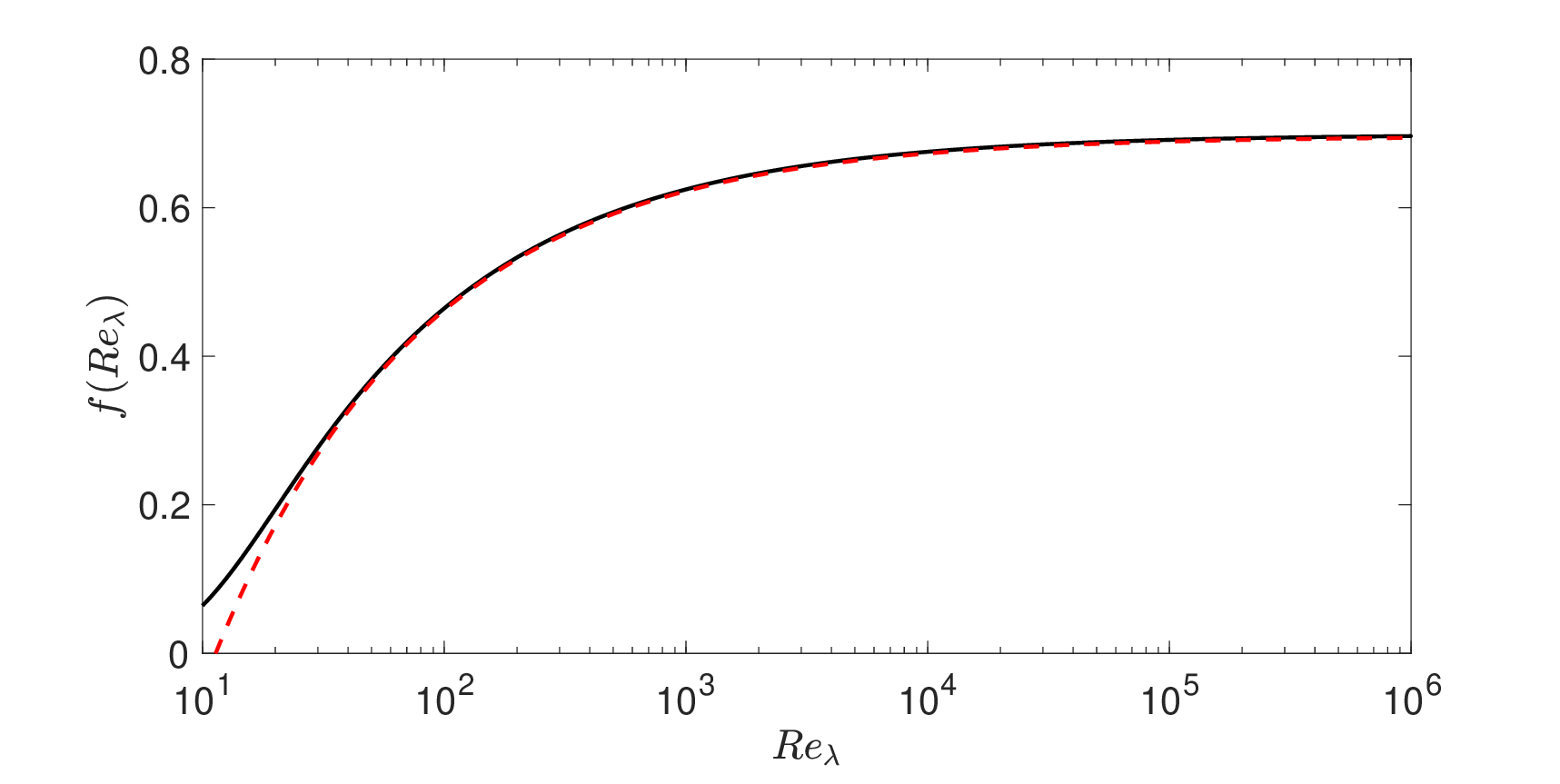}
\caption{Variation of the function $f(Re_{\lambda})$ with $Re_{\lambda}$. The function increases monotonically with $Re_{\lambda}$ and asymptotically approaches the value $0.698$ as $Re_{\lambda} \rightarrow \infty$. The solid line represents the numerical evaluation of \eqref{expresiion_f(Re_lambda)}, while the dashed line corresponds to the asymptotic expression \eqref{asymptotic_fRelam}.}
\label{Relam_vs_fRelam_plot}
\end{figure}

Figure \ref{Relam_vs_fRelam_plot} shows the variation of $f(Re_{\lambda})$ with $Re_{\lambda}$. As $Re_{\lambda}$ increases, $f(Re_{\lambda})$ approaches the asymptotic value $f_{\infty} \approx 0.698$. The numerical evaluation of \eqref{expresiion_f(Re_lambda)} is compared with the asymptotic prediction $f(Re_{\lambda})=f_{\infty} - k Re_{\lambda}^{-1/2}$, and the agreement becomes increasingly good with increasing $Re_{\lambda}$. This comparison supports the asymptotic analysis and indicates that the leading finite-$Re_{\lambda}$ correction is proportional to $Re_{\lambda}^{-1/2}$. The asymptotic expression is therefore expected to be useful for estimating $f(Re_{\lambda})$ at moderately large $Re_{\lambda}$ without evaluating the integral in \eqref{expresiion_f(Re_lambda)} numerically.

\section{Expressions for the coefficients in equation \eqref{Final_ode_for_the_pair_probability}}\label{appC}
The expressions for the coefficients appearing in \eqref{Final_ode_for_the_pair_probability} are given by 
\begin{align}
   \mathcal{G}_n^m(r) &= \left(A-1\right)^2\Big\{\frac{24}{35}\mathcal{W}_{4nm} + \frac{8}{21}\mathcal{W}_{2nm} - \frac{32}{15\left(2n+1\right)}\delta_{nm}\Big\},\\
   \mathcal{H}_n^m(r) &= \left(A-1\right)\Big[\Big\{\frac{24\left((2n-3)(B-1)+X\right)}{35}\mathcal{W}_{4nm} \nonumber \\ & + \frac{(44n+32)B+12(n-5)+8X}{21}\mathcal{W}_{2nm} \nonumber \\ & + \frac{4(n+1)(2B+3)-16X}{15}\frac{2}{\left(2n+1\right)}\delta_{nm}\Big\} \nonumber \\ & - (n+1)\Big\{\frac{12\left(B-1\right)}{5} \mathcal{W}_{3(n+1)m} + \frac{4\left(2B+3\right)}{5}\mathcal{W}_{1(n+1)m}\Big\}\Big],\\
   \mathcal{I}_n^m(r) &= L\mathcal{W}_{1nm},\\
   \mathcal{J}_n^m(r) &= n(n+1)\Big\{\frac{24\left(B-1\right)^2}{35}\mathcal{W}_{4nm} - \frac{2B\left(3B-34\right)+62}{21}\mathcal{W}_{2nm} \nonumber \\ & + \frac{B\left(9B-28\right)+34}{15}\frac{2}{\left(2n+1\right)}\delta_{nm}\Big\} + (n+1)\Big\{\frac{24Y}{35}\mathcal{W}_{4nm} \nonumber \\ & + \frac{2\left(27Y+14Z\right)}{105}\mathcal{W}_{2nm}  +\frac{2Z}{15}\frac{2}{\left(2n+1\right)}\delta_{nm}\Big\} \nonumber \\ & + \frac{24}{35}\Big\{\left(T+r\frac{dT}{dr}\right) + W\left(2A-5B+3\right)\Big\}\mathcal{W}_{4nm} \nonumber \\ & + \frac{1}{21}\Big\{8\left(T+r\frac{dT}{dr}\right) + W\left(16A-12B-88\right)\Big\}\mathcal{W}_{2nm} \nonumber \\ & + \frac{16}{15}\Big\{\left(T+r\frac{dT}{dr}\right) + 2W\left(A-1\right)\Big\}\frac{2}{\left(2n+1\right)}\delta_{nm} \nonumber \\ & -
   (n+1)\Big\{\frac{6Y}{5}\mathcal{W}_{3(n+1)m} + \frac{2Z}{5}\mathcal{W}_{1(n+1)m}\Big\},\\
   \mathcal{N}_n^m(r) &=  \Big\{(n+1)M + 2\left(L-M\right) + r\frac{dL}{dr}\Big\}\mathcal{W}_{1nm} \nonumber \\ & - (n+1)M\frac{2}{\left(2n+3\right)}\delta_{(n+1)m}. 
\end{align}
Here, $\mathcal{W}_{lnm}$ the integral of the product of three Legendre polynomials, which can be expressed in terms of Wigner 3j symbols as
\begin{equation}
    \mathcal{W}_{lnm} = \int_{-1}^1\mathscr{P}_l(\mu)\mathscr{P}_n(\mu)\mathscr{P}_m(\mu)d\mu = 2 \left(
    \begin{array}{ccc}
    l & n & m \\
    0 & 0 & 0
    \end{array} \right)
    \left(
    \begin{array}{ccc}
    l & n & m \\
    0 & 0 & 0
    \end{array} \right).
\end{equation}
The parameters $X$, $Y$, $Z$, and $T$ are functions of the hydrodynamic mobilities $A$ and $B$, and are given by
\begin{align}
    X &= 7A - 3B + 3r\frac{dA}{dr} - 4, \\
    Y &= \frac{6}{5}\Big\{\left(B-1\right)\left(6A-7B+2r\frac{dA}{dr}+1\right) + \left(A-1\right)r\frac{dB}{dr}\Big\}, \\
    Z &= \frac{2}{5}\Big\{\left(2B+3\right)\left(6A-3B+2r\frac{dA}{dr}-3\right) + 2\left(A-1\right)r\frac{dB}{dr} - \left(3B+17\right)\left(B-1\right)\Big\}, \\
    T &= -\left\{3(A-B)+r\frac{dA}{dr}\right\}\left(A-1\right).
\end{align}

\bibliographystyle{jfm}
\bibliography{jfm}

\end{document}